\documentclass[10pt,aps,pra,twocolumn,showpacs,superscriptaddress,nobalancelastpage,notitlepage]{revtex4-1}

\usepackage{graphicx}
\usepackage{amsmath}
\usepackage{longtable}
\usepackage{latexsym}
\usepackage{times}
\usepackage{mathtools}
\usepackage[T1]{fontenc}
\usepackage{float}
\usepackage{enumerate}
\usepackage{setspace,ifthen}
\usepackage{amsthm} % Theorem Formatting
\usepackage{amssymb}% Math symbols such as \mathbb
\usepackage[usenames,dvipsnames]{xcolor}
\usepackage{mathrsfs}
\usepackage{units}
\usepackage{braket}
\usepackage{upgreek}
\usepackage[hidelinks, bookmarks=false]{hyperref}
\usepackage[colorinlistoftodos]{todonotes}

\newcommand{\expect}[1]{\langle #1 \rangle}

\newcommand{\erf}[1]{Eq. (\ref{#1})}

\def\XXint#1#2#3{{\setbox0=\hbox{$#1{#2#3}{\int}$}
     \vcenter{\hbox{$#2#3$}}\kern-.5\wd0}}

\begin{document}

%\title{Optimally band-limited controls for {\color{magenta} simultaneous spectral estimation of dephasing and control noise}}

\title{Simultaneous spectral estimation of dephasing and amplitude noise on a qubit sensor \\ 
via optimally band-limited control}

\date{\today}

\author{Virginia Frey}
\affiliation{ARC Centre for Engineered Quantum Systems, School of Physics, The University of Sydney, NSW 2006 Australia}

\author{Leigh M. Norris}
\affiliation{\mbox{Department of Physics and Astronomy, Dartmouth 
College, 6127 Wilder Laboratory, Hanover, NH 03755, USA}}

\author{Lorenza Viola}
\email{Contact: lorenza.viola@dartmouth.edu}
\affiliation{\mbox{Department of Physics and Astronomy, Dartmouth 
College, 6127 Wilder Laboratory, Hanover, NH 03755, USA}}

\author{Michael J. Biercuk}
\email{Contact: michael.biercuk@sydney.edu.au}
\affiliation{ARC Centre for Engineered Quantum Systems, School of Physics, The University of Sydney, NSW 2006 Australia}

\begin{abstract}
The fragility of quantum systems makes them ideally suited for sensing applications at the nanoscale. However, interpreting the output signal of a qubit-based sensor is generally complicated by background clutter due to out-of-band spectral leakage, as well as ambiguity in signal origin when the sensor is operated with noisy hardware. We present a sensing protocol based on optimally band-limited ``Slepian functions'' that can overcome these challenges, by providing narrowband sensing of ambient dephasing noise, coupling additively to the sensor along the ${z}$-axis, while permitting isolation of the target noise spectrum from other contributions coupling along a different axis. This is achieved by introducing a finite-difference control modulation, which linearizes the sensor's response and affords tunable band-limited ``windowing'' in frequency.  Building on these techniques, we experimentally demonstrate two spectral estimation capabilities using a trapped-ion qubit sensor. We first perform efficient experimental reconstruction of a ``mixed'' dephasing spectrum, composed of a broadband $1/f$-type spectrum with discrete spurs.  We then demonstrate the simultaneous reconstruction of overlapping dephasing and control noise spectra from a single set of measurements, in a setting where the two noise sources contribute equally to the sensor's response. Our approach provides a direct means to augment quantum-sensor performance in the presence of both complex broadband noise environments and imperfect control signals, by optimally complying with realistic time-bandwidth constraints.
\end{abstract}

\maketitle
%\tableofcontents

\section{Introduction}

Quantum sensors harness a feature which is otherwise regarded as the central weakness of quantum technologies as a resource: their extreme sensitivity to external disturbances. Applications range from magnetometry and medical imaging to noise characterization for optimized control design in intermediate-scale quantum computers and simulators \cite{Degen2017, Alvarez_Spectroscopy, Bylander2011}. In conventional operation, and in the simplest setting where a single qubit is employed as a sensor, the sensor undergoes free evolution during which an integrated signal from the environment changes the qubit's state in a measurable way~\cite{Degen2017}.  This form of ``Ramsey experiment'' exhibits broad-band coupling to the environment with sensitivity down to DC.

Adding time-dependent control to the sensor provides a means to adjust its spectral response -- as is needed for applications in frequency-tuned sensing~\cite{taylor2008high,Retzker2017,Kosuke2018}. This general approach has been employed in dynamical-decoupling noise spectroscopy protocols in either pulsed~\cite{Bylander2011,Alvarez_Spectroscopy,Norris_Spectroscopy} or continuously-driven form~\cite{Hirose12}, as well as in spin-locking-based protocols \cite{Yan2013,Yan2018}.  However, existing spectral estimation approaches leveraging such protocols suffer from significant drawbacks. First, while pulsed protocols based on frequency-comb sampling have been theoretically extended to estimation of multi-axis additive noise \cite{PazNorris2019}, they do not account for control noise and involve abrupt transitions in the amplitude or phase of the applied drive \cite{Carr1954,Meiboom1958,Viola1999,Cappellaro2013,Ball2015}, which inevitably result in additional sensitivity outside of the target frequency band. This phenomenon, known as \emph{spectral leakage}, can cause ambiguity in the interpretation of the sensor response, as out-of-band signals can couple to harmonics of the target band induced by the rapid control transitions~\cite{degen_retract,degen_followup}. Second, any imperfections on the control itself, or contributions from other unwanted Hamiltonian terms, are manifested as deviations in the qubit-sensor's state that are indistinguishable from the target signal in conventional projective measurements. 

Here, we present a continuously driven, smoothly modulated control protocol for qubit sensors, which employs optimally band-limited Slepian functions, more formally known as \emph{discrete prolate spheroidal sequences} (DPSS)~\cite{Slepian1983}. Widely used in classical statistical signal processing \cite{percival1993spectral}, DPSS have recently found application in optimal control algorithms for quantum gate synthesis \cite{Lucarelli2017} and enabled a proof-of-concept demonstration of multitaper spectral estimation in the limited setting of multiplicative noise on the driving amplitude -- collinear (hence commuting) with the control axis  ($\propto\sigma_x$) \cite{Frey2017, Norris2018}. While this type of noise commonly arises from the control hardware in platforms ranging from trapped ions to solid-state qubits \cite{Dial2013,Didier2019}, an even more prevalent (or concomitant) form of noise is control-independent, dephasing noise that couples in an additive fashion along the quantization axis of the qubit ($\propto\sigma_z$). Extending DPSS-based spectral estimation to include non-commuting additive dephasing noise, while maintaining the desired spectral concentration in the frequency domain, requires introducing a qualitatively different control modulation, able to linearize the sensor's response and effectively invert the ensuing non-linearity via a finite-difference scheme. 

We show that our approach for synthesizing optimally concentrated filters, coupled with tomographic measurement of the sensor's state, provides simultaneous, tunable, narrowband responses to \emph{both} the non-commuting dephasing signal and the commuting, multiplicative noise terms. We experimentally demonstrate the efficacy of these controls using a single trapped $^{171}$Yb$^{+}$ ion, by mapping the filter function~\cite{Kofman2004,GreenNJP2013, Ball2014, PazFFF} of the control in multiple Cartesian projections.  We then showcase the ability to reconstruct an engineered ``mixed'' dephasing spectrum \cite{percival1993spectral}, composed of both broadband and narrowband features, through a Bayesian estimation procedure.  Finally, we leverage the narrowband properties of our controls along multiple Cartesian projections to \emph{simultaneously} reconstruct two overlapping spectra, arising from noise in the amplitude ($\propto\sigma_x$) and phase ($\propto\sigma_z$) quadratures, using the \emph{same} set of tomographic measurements.

%% LV: Do we need a content summary here?
%% The remaining content is organized as follows.  ...brief summary here...}

%%%%% Main body 

%\section{Designing spectrally concentrated filters for quantum sensing}
%\subsection*{\textcolor{magenta}{Background}: System and control setting} 
%% LV: I think that it may be better to separate Background (= known from previous work) from 
%%       Finite-difference, which from the theory standpoint is the key innovation of this work

\section{Background}
\label{sec::back}

\subsection{System and control setting}
\label{sub::setting}

\noindent
In previous work~\cite{Frey2017, Norris2018}, we showed how qubit sensors controlled with DPSS-shaped waveforms possess a narrowband frequency response to a target noise term that commutes with the applied control -- referred to here as ``control noise''. We now expand on this by including a non-commuting, dephasing noise term, and present a protocol that has the ability to reconstruct \emph{both} the control noise spectrum as well as the dephasing noise spectrum \emph{simultaneously}. As before, we consider a qubit sensor that is subject to external control and time-dependent noise. The action of control in the rotating frame is captured by the following Hamiltonian ($\hbar=1$),
\begin{equation}
H_{\text{ctrl}} (t) = \frac{\Omega(t)}{2} (\cos \varphi (t) \sigma_x + \sin \varphi (t) \sigma_y),
\label{eq:ctrl}
\end{equation}
where $\Omega(t)$ is the time-dependent amplitude of the driving field, which is tunable within a maximum range $|\Omega|\leq \Omega_{\text{max}}$, and $\varphi(t)$ is a time-dependent control phase. Since in this work we primarily employ single-axis controls, we let $\varphi(t) = 0$, yielding the ideal control Hamiltonian $H_\text{ctrl} (t) \equiv \Omega(t)\sigma_x/2$. 

We model the effect of noise through an additive term that represents ambient dephasing noise, and a multiplicative term proportional to the drive amplitude that describes control noise. The corresponding stochastic Hamiltonian reads
\begin{equation}
H_{\text{N}} (t) = \beta_z(t)\sigma_z  + \beta_\Omega (t) H_\text{ctrl}(t) ,
\end{equation}
where $\beta_\Omega(t)$ and $\beta_z(t)$ are independent, stationary and zero-mean Gaussian processes.  Physically, this model Hamiltonian provides an accurate description for a variety of qubit sensors operating in a classical noise regime, with dephasing ($T_2$) processes resulting, for instance, from a semiclassical treatment of spin or bosonic environments \cite{Witzel2014,Degen2017,Szankowski2017}, and multiplicative noise arising from hardware noise -- for instance, fluctuations in the applied microwave power in our trapped-ion setting, dc voltage fluctuations in exchange-controlled spin qubits \cite{Dial2013}, or ac voltage fluctuations in superconducting qubits \cite{Didier2019}. Under the Gaussian assumption, the noise properties are fully characterized in terms of their power spectral densities, obtained through the Fourier transforms of the respective auto-correlation function, 
$${S_{u}(\omega) = \frac{1}{2\pi} \int_{-\infty}^{\infty} \mathrm{d}s \langle \beta_{u}(0) \beta_{u}(s)\rangle \mathrm{e}^{-\mathrm{i}s \omega}}, \quad u \in \{\Omega, z \}.$$

The total Hamiltonian of a qubit sensor undergoing both driven control and noise is then given by
\begin{equation}
H(t) = \beta_{z}(t)\sigma_z + \Omega(t) [1 + \beta_{\Omega}(t)]\sigma_{x}/2.
\label{eq:basic_hamiltonian}
\end{equation}
%\begin{figure*}[t!]
%\includegraphics[width=\textwidth]{slepian_introduction_figure.pdf}
%\label{fig:slepian_introduction}
%\caption{\textcolor{magenta}{The DPSS and their application in noise sensing. \textbf{a} shows four sample DPSS of orders $k=0$ to $k=3$ with $N=500$ points and time step $\Delta t = \unit[2]{\upmu s}$ (total duration of $\unit[1]{ms}$) and time-bandwidth product $NW=4$. \textbf{b} shows the absolute magnitude of their respective Fourier transforms, the DPSWFs. The shaded area indicates the band over which they exhibit optimal spectral concentration. \textbf{c} illustrates how much spectral leakage can be suppressed when using a $k=0$th-order DPSS instead of a traditional, flat-top control pulse. The suppressed leakage enables more accurate spectral estimation, as panel \textbf{d} demonstrates.}}
%\end{figure*}

It is further convenient to effect a transformation to a frame that co-rotates with the ideal control, that is, to consider $\widetilde{\rho}(t)\equiv U_{\text{ctrl}}^\dagger (t)\rho(t) U_{\text{ctrl}}(t)$, where $U_{\text{ctrl}}(t)$ is the propagator generated by the ideal control Hamiltonian $H_{\text{ctrl}} (t)$ and  $\rho(t)$ is the qubit density operator in the physical (rotating) frame. The Hamiltonian in Eq. \eqref{eq:basic_hamiltonian} then maps to
$$\widetilde{H}(t)=[\cos \Theta(t) \sigma_z + \sin\Theta(t) \sigma_y] \beta_{z}(t) + \Omega(t)\beta_{\Omega}(t) \sigma_{x}/2, $$ where the integrated angle of driven rotation is \mbox{$\Theta(t) \equiv \int_0^t \mathrm{d}s \Omega(s)$}. In this frame, the time evolution over an interval $[0, \tau]$ is described by the unitary propagator
\begin{equation}
  \widetilde{U}(\tau) = \mathcal{T}_+ \exp\left[ \int_0^\tau \mathrm{d}s \widetilde{H}(s) \right] \equiv \exp[- \mathrm{i} \mathbf{a}(\tau)\cdot \boldsymbol{\sigma}],
  \label{eq:u_tilde}
\end{equation}
\noindent

which is related to the rotating-frame propagator $U(t)$ via $\widetilde{U}(t)= U^\dagger_{\text{ctrl}}(t) U(t)$. Here,
$\boldsymbol{\sigma}$ is the Pauli vector and we have defined \mbox{$\mathbf{a}(\tau) \equiv [a_x(\tau), a_y(\tau), a_z(\tau)]$} as a real, time-dependent (stochastic) ``error vector''~\cite{GreenNJP2013} that captures the evolution due to noise.  We have chosen a Cartesian representation, as it maps to standard tomographic protocols for measuring qubit state projections. Throughout our analysis and experiments, we shall work in a regime where the noise is sufficiently weak and the time scales are sufficiently small, such that we only need to consider the leading (first) order terms in a perturbative Magnus expansion of the error vector \cite{SoareNatPhys2014, Ball2015, Norris2018}, that is, $\mathbf{a}(\tau) \approx \mathbf{a}^{(1)}(\tau)$. Explicitly, we have:  
\begin{subequations}
\label{eq:error_vector_components_time}
\begin{align}
    a_x^{(1)}(\tau) &= \frac{1}{2} \int_0^\tau \mathrm{d}s \,\Omega(s)\beta_{\Omega}(s) ,
    \label{eq:error_vector_components_time_x}\\
    a_y^{(1)}(\tau) &= \int_0^\tau \mathrm{d}s \sin\Theta(s)\beta_{z}(s) ,
    \label{eq:error_vector_components_time_y}\\
    a_z^{(1)}(\tau) &= \int_0^\tau \mathrm{d}s \cos\Theta(s)\beta_{z}(s).
\label{eq:error_vector_components_time_z}
\end{align}
\end{subequations}
Here the control noise, $\beta_{\Omega}(t)$, enters $a_x^{(1)}$ in a way that is linearly proportional to the control variable, $\Omega(t)$. In contrast, the dephasing noise, $\beta_{z}(t)$, enters both $a_y^{(1)}$ and $a_z^{(1)}$ through the nonlinear function $\Theta(\tau)$, which in turn causes the dephasing to couple in a highly nonlinear way to $\Omega(t)$.
%and this dephasing term thus couples to the applied control in a nonlinear way
%, which is the main reason for why qubit sensors are commonly driven with pulsed instead of continuous control (see Fig. \ref{fig:finite_diff_filters}b-c). In this work, we introduce a method to \emph{linearize} this dephasing response to the applied control, thereby enabling the use of arbitrary, continuous waveforms for dephasing noise spectroscopy.}
%% LV: Virginia, I do not think this is very helpful here - I do not understand what you mean by 'the main reason'... and, referring to Fig.1 here comes way too soon to make sense w/o more explanation...

The above error vector components can be accessed in the experiment through a ``three-axis measurement protocol'' of the qubit state \cite{Frey2017}. If we prepare the qubit in the positive $z$-state, denoted here as $|\uparrow_z\rangle$, with $\sigma_z |\uparrow_z\rangle = |\uparrow_z\rangle$, the survival probability in this state under evolution of $\widetilde{U}(\tau)$ from Eq. \eqref{eq:u_tilde}, that is, $P(\uparrow_z) = |\langle \uparrow_z | \widetilde{U}(\tau) | \uparrow_z \rangle|^2$, is approximately given by
\begin{align}
P(\uparrow_z)  \approx 1 - \langle |a_x^{(1)}(\tau) |^2\rangle - \langle |a_y^{(1)}(\tau) |^2\rangle,
\label{eq:three_axis_p_z}
\end{align}
 where $\langle \:,\:\rangle$ denotes the ensemble average taken over all possible time-domain realizations of the stochastic process, and we have used the same first-order truncation assumed in Eq. \eqref{eq:error_vector_components_time}. Equivalently, when preparing the qubit in the states $|\uparrow_x \rangle$ and $| \uparrow_y \rangle$, we find for $P(\uparrow_x)$ and $P(\uparrow_y)$
\begin{align}
  P(\uparrow_x) &\approx 1 - \langle |a_y^{(1)}(\tau) |^2\rangle - \langle |a_z^{(1)}(\tau) |^2\rangle, \label{eq:three_axis_p_x}\\
  P(\uparrow_y) &\approx 1 - \langle |a_x^{(1)}(\tau) |^2\rangle - \langle |a_z^{(1)}(\tau) |^2\rangle.\label{eq:three_axis_p_y}
\end{align}
The individual error vector components can then be reconstructed by taking appropriate linear combinations of $P(\uparrow_x)$, $P(\uparrow_y)$ and $P(\uparrow_z)$.

\begin{figure*}[t!]
\centering
\includegraphics[scale=1]{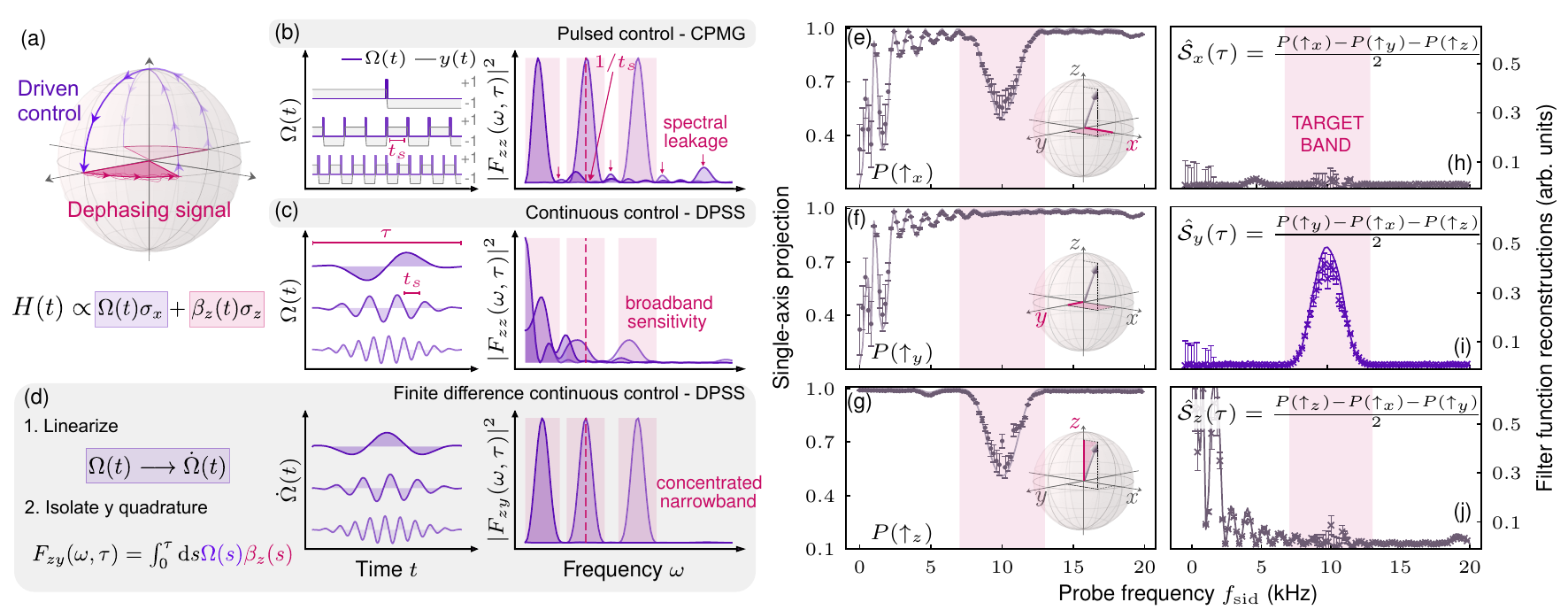}
\caption{Designing spectrally concentrated filters for dephasing sensing. \textbf{(a)} illustrates the dephasing sensing process  (Hamiltonian in Eq. \eqref{eq:basic_hamiltonian} with $\beta_\Omega(t) \equiv 0$), involving periods of phase accumulation under the dephasing process $\beta_z(t)$ and driven control $\Omega(t)$. \textbf{(b, c)} show  $\Omega(t)$ for conventional sensing sequences and the corresponding FFs for both pulsed (CPMG) and continuous (DPSS) modulation. The FF for pulsed control, $|F_{zz}(\omega, \tau)|^2=|\int_0^\tau \mathrm{d}s\,e^{i\omega s}y(s)|^2$, exhibits harmonics at $\omega=\pi n/t_s$ for integer $n \ge 1$. For the DPSS modulation in \textbf{(c)}, $F_z(\omega,\tau)= |F_{zy}(\omega, \tau)|^2$ is broadband sensitive. \textbf{(d)} Finite-difference controls linearize the sensor response to the target dephasing signal. The corresponding FFs are tunable and spectrally concentrated. \textbf{(e, f, g)} Experiments show the response of a finite-difference control waveform built from a zeroth-order DPSS to a range of probe frequencies, $f_{\mathrm{sid}}$, applied in the form of a single-frequency modulation in the dephasing quadrature. Measurements are taken in the $x$, $y$ and $z$ quadrature as the survival probability along the corresponding Bloch sphere projection -- see Eqs. (\ref{eq:three_axis_p_z})--(\ref{eq:three_axis_p_y}) in the main text. Experimental data is represented by markers, while continuous lines show numerical simulations. Each data point comprises an average over 500 individual repetitions of the experiment and the error bars represent the standard deviations of those averages. \textbf{(i, j)} show the estimated dephasing filters for control duration $\tau$, $\hat{\mathcal S}_i(\tau) =  \langle |\hat{a}_i^{(1)}(\tau) |^2\rangle \approx |\hat{F}_{zi}(\omega_{\mathrm{sid}},\tau)|^2$, $i \in \{y, z\}$, calculated from linear combinations of the projection data (inset equations). \textbf{(h)} consistently shows no sensitivity to dephasing noise in the signal's $x$ projection, ${\mathcal S}_x (\tau) =  \langle |a_x^{(1)}(\tau) |^2\rangle \approx |\hat{F}_{xx}(\omega_{\mathrm{sid}},\tau)|^2$. In \textbf{(i)} we recover the desired spectrally concentrated dephasing filter $F_{z}(\omega, \tau)$. }
\label{fig:finite_diff_filters}
\end{figure*}

Moving to the frequency domain, the action of the external control is most conveniently described within the filter function (FF) formalism~\cite{GreenNJP2013, PazFFF, KavehDCG, SoareNatPhys2014}. By using the explicit form of $H_{\mathrm{ctrl}} (t)$, three fundamental FFs suffice to evaluate how the sensor's response to dephasing and amplitude noise is modified by the control \cite{Norris2018}, namely, the Fourier transforms 
\begin{eqnarray}
\label{eq:fff}
F_{xx}(\omega, \tau) & \equiv &\int_0^\tau  \mathrm{d} s \,\Omega(s) e^{i\omega s} , 
\label{fxx}\\
F_{zy}(\omega, \tau) & \equiv &\int_0^\tau  \mathrm{d}s \sin \Theta(s) e^{i\omega s} , 
\label{fzy}\\
F_{zz}(\omega, \tau) & \equiv &\int_0^\tau  \mathrm{d}\tau \cos \Theta(\tau) e^{i\omega\tau} .
\label{fzz}
\end{eqnarray}
The error vector components from Eq.~\eqref{eq:error_vector_components_time} may then be expressed as overlap integrals between appropriate FFs and the corresponding noise spectra, 
\begin{subequations}
\label{eq:axy}
\begin{align}
\langle |a_x^{(1)}(\tau) |^2\rangle & = \frac{1}{\pi}\int_0^\infty  \mathrm{d} \omega F_\Omega(\omega, \tau) S_\Omega(\omega) , 
\label{ax}\\
\langle |a_y^{(1)}(\tau) |^2\rangle & = \frac{1}{\pi}\int_0^\infty  \mathrm{d} \omega F_z (\omega, \tau) S_z(\omega) , 
\label{ay}
\end{align}
\end{subequations}
where the \emph{amplitude FF} and \emph{dephasing FF} are given by
\begin{subequations}
\begin{align}
F_\Omega(\omega, \tau) &\equiv \frac{1}{4} |F_{xx}(\omega, \tau)|^2, \label{eq::amplitude_FF} \\
F_z (\omega, \tau) &\equiv |F_{zy}(\omega, \tau)|^2. \label{eq::dephasing_FF}
\end{align}
\end{subequations}

The primary tool available for shaping these filters and thus changing the sensor's spectral response is temporal modulation of the control amplitude, $\Omega(t)$. Common control protocols like, for instance, continuously driven rotary spin echoes provide a tunable response to a target control noise spectrum \cite{Frey2017}, while discretely pulsed sequences like the Carr-Purcell-Meiboom-Gill (CPMG) sequences are known to provide a tunable response of the sensor to a target dephasing spectrum (see Fig. \ref{fig:finite_diff_filters}b). In the latter case, when taking the limit of instantaneous $\pi_x$ pulses, the control-dependent term in Eq. (\ref{eq:error_vector_components_time_z}) becomes a piecewise-constant function $y(t)$ that switches between $\pm 1$ whenever a pulse is applied. The corresponding FF, given by the Fourier transform of this rectangular switching function, then has a sinc-like shape with infinite harmonics at integer multiples of the pulse separation.

\medskip

\subsection{DPSS control modulation}

\noindent
Continuous modulation of $\Omega(t)$ can be employed to reduce spectral leakage associated with such sharp transitions. In our previous work, we demonstrated how DPSS-shaped control waveforms produce optimally spectrally concentrated $F_{xx}(\omega, \tau)$ FFs, which allow for minimally biased spectral estimation of control noise spectra. To briefly recap, the DPSS (see Fig. \ref{fig:finite_diff_filters}c) are formally defined as the solutions to the eigenvalue problem,
\begin{equation*}
  \sum_{m=0}^{N-1} \!\frac{\sin 2\pi W(n-m)}{\pi(n-m)} \, v_m^{(k)} (N,W) = \lambda_k(N, W) v_n^{(k)}(N, W), 
\end{equation*}
where $N$ is the number of points in the time-domain, $W$ the half-bandwidth parameter, and $v_n^{(k)} (N,W)$ the $n$th element of the $k$th-order DPSS with $n, k \in \{0, 1,\dots, N\!-\!1 \}$. The eigenvalue $\lambda_k(N, W)$ determines the extent of the \emph{spectral concentration} of the DPSS in the frequency band $B_0\equiv (-2\pi W/\Delta t,2\pi W/\Delta t)$. Specifically, taking the discrete Fourier transforms of the DPSS results in the so-called discrete prolate spheroidal waveforms (DPSWF), which read
\begin{align}
 \label{eq::DPSWF}
 \!U^{(k)} \!(N,W;\omega)\equiv\epsilon_k\!\sum_{n=0}^{N-1}\!v_n^{(k)}\!(N,W)e^{i\omega[n-(N-1)/2]\Delta t},
 \end{align}
where $\epsilon_k=1\,(i)$ for even (odd) $k$, respectively. Crucially, the DPSWF $U^{(k)} \!(N,W;\omega)$ provide \emph{optimal spectral concentration}, in the sense that they provably maximize the ratio of the ``signal energy'' (as quantified by the integral of $|U^{(k)} \!(N,W;\omega)|^2=U^{(k)} \!(N,W;\omega)^2\,$) in the target passband $B_0$ to that in the principal domain ($-\pi/\Delta t, \pi/\Delta t$).

To take advantage of these properties in our qubit control, the simplest approach, employed in \cite{Frey2017}, is 
to design an amplitude control waveform divided into $N$ piecewise-constant increments of duration $\Delta t$, 
\begin{align}
\Omega_\text{DPSS}(t)=\Omega\, v_n^{(k)}(N,W),\quad t\in[n\Delta t,(n+1)\Delta t),
 \label{eq::SuppDPSSMod}
\end{align}
where $\Omega$ is a constant scaling factor in units of frequency and the total duration $\tau=N\Delta t$.  Using \erf{eq::DPSWF} together with \erf{eq::amplitude_FF}, we see that the basic DPSS control modulation described above produces the amplitude FF
 \begin{align*}
  F_\Omega(\omega,\tau)=&\,\frac{\Omega^2\sin^2(\omega\Delta t/2)}{\omega^2}\,U^{(k)}(N,W;\omega)^2.
\end{align*}
This FF inherits the spectral concentration properties of $U^{(k)}(N,W;\omega)$ and is therefore also concentrated in the band $B_0$, which by construction is centered around $\omega=0$. In fact, these concentration properties allow for out-of-band leakage suppression of up to $\unit[80]{dB}$ compared to traditional pulsed control~\cite{Frey2017}. For frequency-selective sensing, this approximate bandpass filter can be ``shifted'' in the frequency domain through signal processing techniques such as single-sideband or co-sinusoidal modulation \cite{Frey2017,Norris2018}. To shift the FF by a frequency $\omega_s\geq 0$ using cosinusoidal modulation, the amplitude waveform in \erf{eq::SuppDPSSMod} is modified by
  \begin{align}
  \Omega_\text{COS}(t)=\Omega\cos(n\omega_s\Delta t) &\,v_n^{(k)}(N,W).\;\;
  \label{eq::SuppDPSSMod_2}
  \end{align}
  In the positive half of the frequency domain, the resulting FF is then spectrally concentrated about $\omega_s>2\pi W/\Delta t$,
  \begin{align}
  F_\Omega(\omega,\tau)\approx\,&\frac{\Omega_s^2\sin^2(\omega\Delta t/2)}{2\omega^2}\,\Big [U^{(k)}(N,W;\omega-\omega_s)^2 \nonumber \\
  &+U^{(k)}(N,W;\omega+\omega_s)^2\Big],
  \label{eq::SuppFOmSHift}
  \end{align}
  where the scaling factor $\Omega_s$ is chosen so that the integral $\int_0^T d\mathrm{s}\,\Omega_\text{COS}(s)^2$ is the same for all $\omega_s$. When $\omega_s\leq2\pi W/\Delta t$, the filter has an additional cross-term that can be eliminated through alternative modulation techniques, as discussed in Ref. \cite{Norris2018}. This new band center frequency can be scanned for values of $\omega_s$ up until the maximum sampling bandwidth of the experimental control waveform generator. Therefore, through the DPPS modulation we obtain a tunable, optimally spectrally concentrated filter for control noise sensing.

The general approach outlined above is, however, not directly applicable to the treatment of additive dephasing noise entering the Hamiltonian in \erf{eq:basic_hamiltonian}.  Despite their desirable spectral concentration properties, DPSS-shaped controls suffer from the non-linear dependence on $\Omega(t)$ in the integrand of \erf{fzy}, preventing the DPSWF from emerging in the corresponding FF. Generally, the dephasing FF under simple DPSS modulation of the form given in Eq. \eqref{eq::SuppDPSSMod} is {\em not} spectrally concentrated and thus not suited for ``nonparametric'' estimation \cite{percival1993spectral}, where no \emph{a priori} knowledge about the target signal is assumed. This nonlinearity breaks not only the spectral concentration of the frequency response, 
but also the ability to spectrally tune the sensor when driven with continuous modulation, which is illustrated in Fig. \ref{fig:finite_diff_filters}c: comparing the dephasing FFs of CPMG vs. DPSS controls reveals a breakdown in the peaked sensor response for the latter, irrespective of residual spectral leakage for pulsed control.

\medskip

\section{Designing spectrally concentrated filters via finite-difference}

\noindent
To overcome this fundamental limitation, we return to an analysis of the sensor's noise admittance under these controls. The challenge is to devise a smoothly varying time-domain control modulation $\Omega(t)$, which guarantees a spectrally concentrated response to the dephasing signal, $\beta_z(t)$. We target the dephasing sensitivity of the $y$ error-vector component, given in  Eq.~\eqref{eq:error_vector_components_time_y}, and the corresponding signal projection ${\mathcal S}_y (t) \equiv  \langle |a_y^{(1)}(t) |^2\rangle$, given in Eq. \eqref{ay}. Since the nonlinearity arises from both the sinusoidal dependence and the time-integral over $\Omega(t)$, this may be accomplished by linearizing the sine term and then by compensating the integral term through ``derivative control''.  Under the assumption that the net rotation angle is sufficiently small, $\Theta(t)\ll\pi/2$, we can linearize the sine function in \erf{fzy} to find
\begin{align}
F_{zy}(\omega,\tau)
\approx&\int_0^\tau\!\!ds\,e^{i\omega s}\Theta(s).
\label{eq::SuppFzyLinear}
\end{align}
Since $\Theta(t)=\int_0^tds\,\Omega(s)$, we can create a spectrally concentrated dephasing FF, taking a form similar to \erf{eq::SuppFOmSHift}, by letting $\Omega(t)\propto\frac{d}{dt}\Omega_\text{COS}(t)$. As $\Omega_\text{COS}(t)$ is piecewise constant, however, this derivative does not exist. Instead, we use a waveform depending on the finite difference of $\Omega_\text{COS}(t)$, the discrete analogue of a continuous derivative. If $V_n\equiv\Omega\cos(n\omega_s\Delta t)v_n^{(k)}(N,W)$ denotes the piecewise-constant increments of $\Omega_\text{COS}(s)$, the finite-difference waveform is
\begin{align}
\Omega_\text{FD}(t) =\begin{cases}
V_0'\equiv V_0, & t\in\Delta t\,[0,1)\\
V_1'\equiv V_1\!-\!V_0, & t\in\Delta t\,[1, 2)\\
V_2'\equiv V_2\!-\!V_1, & t\in\Delta t\,[2, 3)\\
\;\;\;\;\;\;\;\;\vdots&\;\;\;\;\;\;\vdots\\
V_{N-1}'\equiv V_{N-1}\!-\!V_{N-2}, & t\in\Delta t\,[N\!-\!2, N\!-\!1).
\end{cases}
\label{eq::SuppFD}
\end{align}
Under this amplitude modulation, observe that the rotation angle at $t=m\Delta t$ becomes
\begin{align*}
\Theta(m\Delta t)=&\int_0^{m\Delta t} \!\!\! ds\,\Omega_{FD}(s) = \sum_{n=0}^{m-1}V_n'\,\Delta t \\ 
=&\,\Omega\Delta t\cos[(m-1)\omega_s\Delta t]v_{m-1}^{(k)}(N,W).
\end{align*}
At each time increment, the rotation angle is thus proportional to $\Omega_\text{COS}(t)$, as desired. For all other times which are not a multiple of $\Delta t$,  the rotation angle is
\begin{align*}
\Theta_\text{FD}(t) =& \int_{t_0}^t \!\! ds\,\Omega_\text{FD}(s)
+ {\sf H} (t_0-1) \!\!\sum_{k=0}^{t_0-1}
\int_{k\Delta t}^{(k+1)\Delta t}\!\!ds\,\Omega_\text{FD}'(s)\notag
\\=&\;V_{t_0}' (t-t_0\Delta t) + {\sf H}\big(t_0-1) \sum_{k=0}^{t_0-1} V_k'\,\Delta t,\notag
\end{align*}
where $t_0 \equiv \lfloor t/\Delta t\rfloor$ denotes the greatest integer less than or equal to $t/\Delta t$, 
and ${\sf H}(\cdot)$ is the discrete Heaviside step function (that is, ${\sf H}(n)= 0, n<0,$ and ${\sf H}(n)=1, n\geq 0$).
Using the relationship between the $V_m$ and $V_m'$ in \erf{eq::SuppFD}, we obtain
\begin{align*}
\Theta_\text{FD}(t)=V_{t_0}(t- t_0 \Delta t) + {\sf H}( t_0-1) 
V_{t_0-1} [(t_0+1) \Delta t-t].  
\end{align*}
By using this expression, it is possible to explicitly evaluate the relevant FFs, as we outline next.

\medskip

\subsection{Finite-difference dephasing filter function}

\noindent
 Substituting the above expression for $\Theta_\text{FD}(t)$ into \erf{eq::SuppFzyLinear} and discretizing the time integral yields (recall that $\tau=N\Delta t$)
\begin{align*}
F_{zy}(\omega,\tau)=&\sum_{m=0}^{N-1}\int_{m\Delta t}^{(m+1)\Delta t}\!\!dt \,e^{i\omega t}\,\Theta(t) %\\
\end{align*}
\begin{align*}
=&\sum_{m=0}^{N-1}V_m\int_{m\Delta t}^{(m+1)\Delta t}\!\!dt \,e^{i\omega t}(t-m\Delta t)\\
-&\sum_{m=1}^{N-1}V_{m-1}\int_{m\Delta t}^{(m+1)\Delta t}\!\!dt \,e^{i\omega t}[t-(m+1)\Delta t]\\
%=&\sum_{m=0}^{N-1}e^{i\omega m\Delta t}\,V_m\int_0^{\Delta t}\!\!dt\,[t+e^{i\omega \Delta t}(\Delta t-t)]\\
%-&\,e^{i\omega N\Delta t}\,V_{N-1}\int_0^{\Delta t}\!\!dt\,e^{i\omega t}(\Delta t-t)\\
\approx &\;e^{i\omega[(N-1)/2]\Delta t}\,\tilde{V}(\omega)\int_0^{\Delta t}\!\!dt\,e^{i\omega t}[t+e^{i\omega \Delta t}(\Delta t-t)]\\
=&\,-e^{i\omega[(N-1)/2]\Delta t}\,\tilde{V}(\omega)\frac{(e^{i\Delta t\omega}-1)^2}{\omega^2}. 
\end{align*}
Here, the tilde denotes a discrete-time Fourier transform (DTFT), i.e., for a discrete sequence $\{\Omega_n\}$,
$\tilde{\Omega}(\omega)\equiv \sum_{n=0}^{N-1}\Omega_n \, e^{i\omega[n-(N-1)/2]\Delta t}.$ To obtain the approximate equality, we have dropped terms $\mathcal{O}(V_{N-1})$, since $V_{N-1}\approx 0$ when $N\gg 1$ for a typical DPSS sequence. By 
recalling \eqref{eq::dephasing_FF} we have
\begin{align}
F_z(\omega,\tau)&=\frac{16\sin^4(\omega\Delta t/2)}{\omega^4}|\tilde{V}(\omega)|^2\\
& \approx\;\frac{8\,\Omega_s^2\sin^4(\omega\Delta t/2)}{\omega^4}\Big[\, U^{(k)}(N,W;\omega-\omega_s)^2 \notag \\
&\,\,\,\,+U^{(k)}(N,W;\omega+\omega_s)^2\,\Big],\label{eq::SuppFz}
\end{align}
where we have taken $\Omega\mapsto\Omega_s$. Thus, the dephasing FF takes a form similar to the amplitude FF under simple DPSS and cosine modulation [\erf{eq::SuppFOmSHift}]. In the positive half of the frequency domain, this yields an optimally spectrally concentrated dephasing filter about $\omega_s$, as illustrated in Fig.~\ref{fig:finite_diff_filters}d.

\subsection{Finite-difference amplitude filter function}
\label{sub::ampFF}

\noindent
Remarkably, under the above protocol, not only does the dephasing FF recover the desired spectral concentration, but concentration is also retained for the amplitude FF.  This may be seen by evaluating $F_\Omega(\omega,\tau)$ in an analogous way to what was done above. This results in 
\begin{align*}
F_{xx}(\omega,\tau) &=\sum_{m=0}^{N-1}\int_{m\Delta t}^{(m+1)\Delta t}\!\!dt \,e^{i\omega t}\,V_m' \\
&= \Big[V_0+\sum_{m=1}^{N-1}e^{i\omega m\Delta t}(V_m-V_{m-1})\Big]\int_0^{\Delta t}\!\!\!dt\,e^{i\omega t} \\
%&=\Big[e^{i\omega[(N-1)/2]\Delta t}\,(1-e^{i\omega\Delta t})\tilde{V}(\omega)+e^{i\omega N\Delta t}\,V_{N-1}\Big]\int_0^{\Delta t}\!\!dt\,e^{i\omega t}\\
& \approx\,e^{i\omega[(N-1)/2]\Delta t}\,\frac{i(-1+e^{i\Delta t\omega})^2}{\omega}\tilde{V}(\omega), 
\end{align*}
again, up to terms $\mathcal{O}(V_{N-1})$. The amplitude FF  is given by \eqref{eq::amplitude_FF}, which produces 
\begin{align*}
F_\Omega(\omega,\tau)
\approx 
%&\,\frac{4\sin^4(\omega\Delta t/2)}{\omega^2}|\tilde{V}(\omega)|^2 \\%+\mathcal{O}(V_{N-1}) \notag \\%= 
&\,\frac{2\,\Omega_s^2\sin^4(\omega\Delta t/2)}{\omega^2}\Big[\,U^{(k)}(N,W;\omega-\omega_s)^2+ \notag\\ 
& \; U^{(k)}(N,W;\omega+\omega_s)^2\,\Big].
\end{align*}
Therefore, to leading order in $V_{N-1}$, the amplitude FF differs from the dephasing FF in \erf{eq::SuppFz} by a factor of $4/\omega^2$.

Importantly, finite-difference control remains compatible with the analog modulation techniques that are needed to shift the filter passband \cite{Frey2017, Norris2018}. In situations where the weak-noise approximation need not hold, the scheme can be further modified to incorporate dynamical decoupling pulses, so to achieve suppression of unwanted higher-order Magnus terms (see Appendix \ref{app::embed}).

\section{Experimental implementation} %and validation} 

\noindent
We experimentally demonstrate the above protocols using a sensor based on a single $^{171}$Yb$^{+}$ ion in a linear Paul trap. The qubit is realized through the hyperfine splitting of the $^1S_{1/2}$ ground state with a transition frequency $\unit[\sim12.6]{GHz}$ \cite{Frey2017,MavadiaNatComms2017,Soare2014bath}. To drive the transition and implement control along both $\sigma_x$ and $\sigma_y$, we employ a vector signal generator with $I/Q$ modulation to produce a control Hamiltonian as in Eq. \eqref{eq:ctrl}. Here, the driving amplitude $\Omega(t)$ is given by the time-dependent $I/Q$ components as $\Omega(t) = \sqrt{I^2(t) + Q^2(t)}$, and the control phase is given by the angle between them as $\varphi(t) = \tan[Q(t)/I(t)]$ (see also Appendix \ref{app:exp_platform}).

Readout is performed through projective measurements in the  $\{|\!\uparrow_z\rangle, |\downarrow_z\rangle\}$ basis. All control waveforms used here implement a  net identity operation ($U_{\text{ctrl}}(\tau)={\mathbb I}$), such that the net evolution under $\widetilde{U}(\tau)$ is solely due to noise. The three-axis measurement procedure from Eq. \eqref{eq:three_axis_p_z}  then permits the ensemble-averaged error-vector components to be estimated via linear combinations of $P(\uparrow_i,t)$, for \mbox{$i \in \{x, y, z\}$}.

Using this three-axis measurement strategy, we demonstrate the narrowband selectivity of finite-difference DPSS controls by reconstructing the controlled sensor's spectral response. We first choose an appropriately constructed DPSS control which is tuned using cosine modulation to shift the target sensing band to $\unit[10]{kHz}$~\cite{Frey2017}. We then employ frequency-selective system identification $(\mathrm{sid})$ to map out the FFs in the presence of additive engineered dephasing noise. A weak, single-frequency disturbance at $\omega_\mathrm{sid}$ is generated by a separate waveform generator and added to the frequency of the driving field via external frequency modulation.  This creates an effective dephasing noise term $\beta_z(t) \sigma_z$, with $\beta_{z}(t)\propto \cos (\omega_{\mathrm{sid}}t + \phi)$, where the variable phase $\phi$ is sampled linearly over $[0, 2\pi)$, in such a way that averaging over $\phi$ yields \mbox{$S_{z}(\omega) \propto \delta (\omega - \omega_{\mathrm{sid}})$} \cite{Soare2014bath}. Leveraging the relationship between signal's projections and FFs yields 
$$\hat{\mathcal S}_i (\tau) = \langle | \hat{a}^{(1)}_i (\tau)|^2\rangle \approx |\hat{F}_{zi}(\omega_{\mathrm{sid}},\tau)|^2, \quad i\in\{y,z\}, $$ 
 where hat symbols are used to differentiate estimated from actual values. Varying $\omega_{\mathrm{sid}}$ and averaging over $\phi$ for each value of $\omega_{\mathrm{sid}}$ then allows direct reconstruction of the FF vs probe frequency.

Data for all Cartesian projections of the measured sensor state are presented in Fig.~\ref{fig:finite_diff_filters}e-g. We observe that all projections exhibit sensitivity to the system-identification stimulus, and all show structure outside of the shaded region representing the target band. However, on inverting these data to reconstruct the FFs, $\hat{\mathcal S}_y(\tau) \approx \hat{F}_{z}(\omega,\tau)$ reveals spectral concentration in the target band with minimal measured leakage, as intended.  The target band is user-defined and experiments performed with different DPSS orders and band shifts reveal comparable performance. In all cases, data agree well with numerical simulations, and the data appearing in Fig.~\ref{fig:finite_diff_filters}i constitute the key validation of our approach to control design.  

\begin{figure}[t!]
\centering
\includegraphics[scale=1.0]{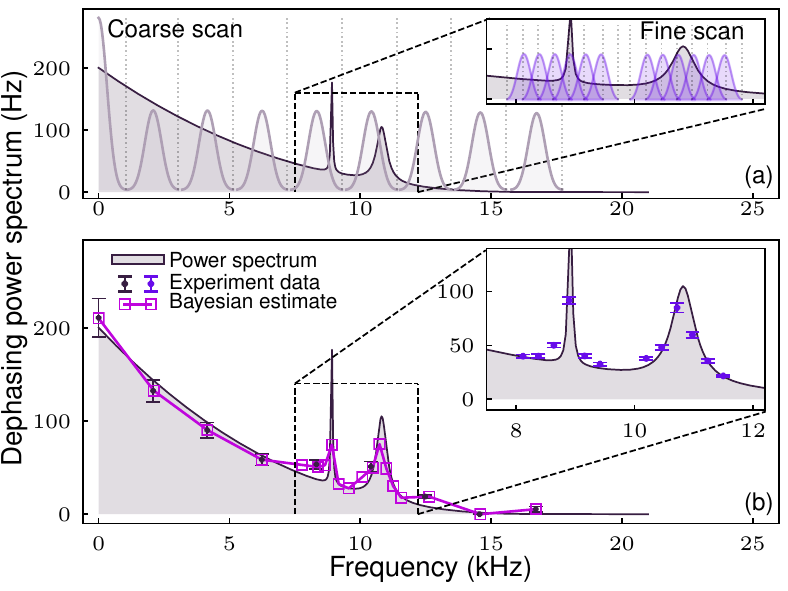}
\vspace*{-4mm}
\caption{Dephasing noise spectrum reconstruction. \textbf{(a)} shows the engineered power spectrum on the left axis, along with the FFs used for the reconstruction. Controls comprised of zeroth-order DPSS with duration $\tau=\unit[2.5]{ms}$, bandwidth product $NW=4$, and band-shift frequencies $f_s =0, 2.1, 4.2, 6.2, 8.3, 10.4, 12.5, 14.6, 16.7$ kHz. The inset shows the location of the filters used in the fine scan, again with zeroth-order DPSS, but now $\unit[5]{ms}$ long with $NW=2$ and $f_s=8.1, 8.4, 8.6, 8.9, 9.2, 9.4,  9.9, 10.2, 10.5, 10.9, 11.2, 11.5$ kHz. \textbf{(b)} shows the experimental spectrum reconstruction of both the coarse (main panel) and fine (inset) scan. Experiments are averaged over 400 time-domain realizations of the noise spectrum and error bars represent the standard deviation
%variance 
%% LV: Virginia, did you mean standard deviation? 
over outcomes. A Bayesian update is employed to combine data from both scans and obtain the final estimate (see also Appendix \ref{app:bayes}).}
\label{fig:dephasing_reconstruction}
\end{figure}

\begin{figure*}[t!]
\centering
\includegraphics[width=\textwidth]{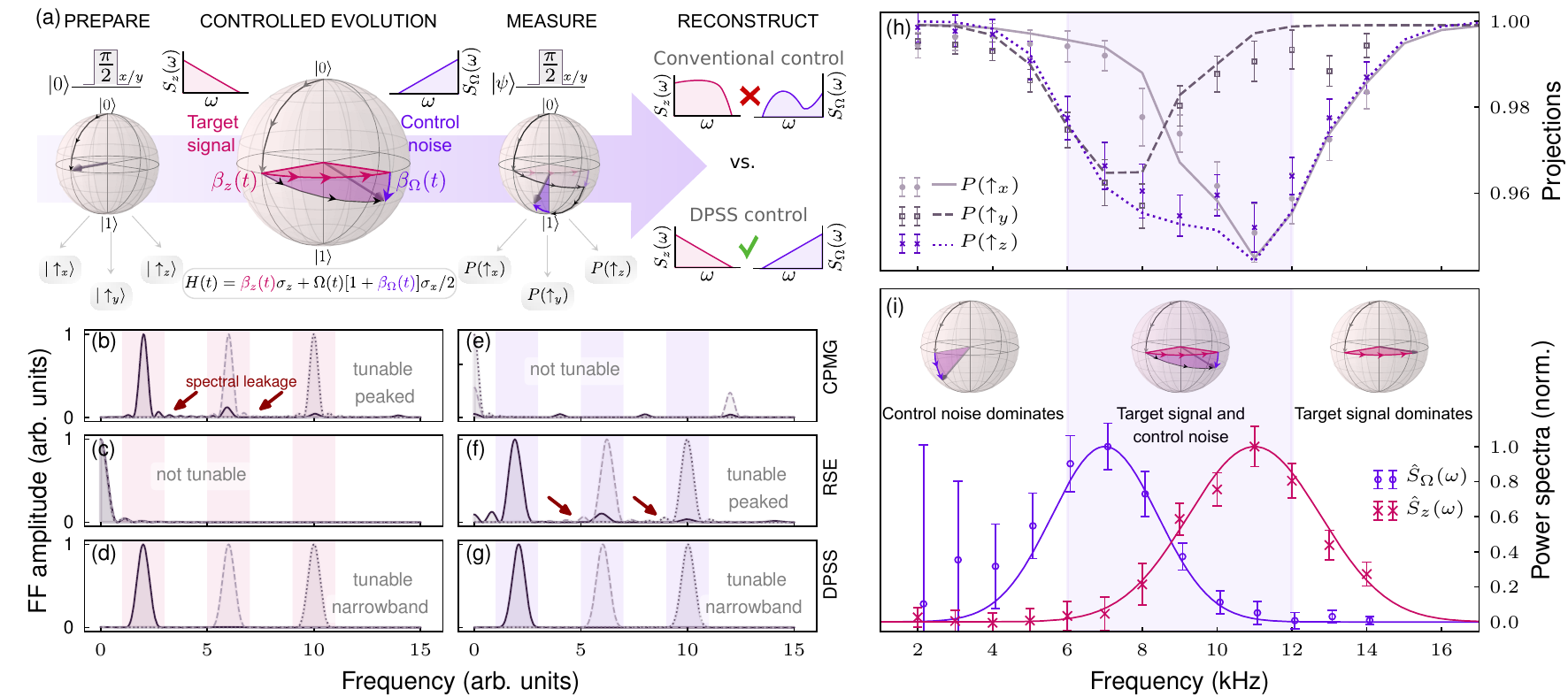}
\caption{Multi-axis sensing of time-dependent signals with a qubit sensor. \textbf{(a)} illustrates the four stages of the sensing process. $\pi/2$ pulses are employed to prepare the qubit in the $+x$, $+y$ and $+z$ state of the Bloch sphere. The qubit is then subjected to a driving field that probes two time-dependent signals, $\beta_{z}(t)$ and $\beta_{\Omega}(t)$, in the target dephasing ($\propto \sigma_z$) and the control ($\propto \sigma_x$) quadrature. Combining measurements in all three axes enables the spectral reconstruction of both signals.  
\textbf{(b, c, d)} show the sensitivity to the target dephasing signal, $\beta_z$, for CPMG control, rotary spin echoes (RSE) and DPSS, by means of the dephasing FFs at three different probe frequencies. \textbf{(e, f, g)} show the corresponding FFs in the control quadrature, as a means to model sensitivity to $\beta_{\Omega}$. Filters have been scaled to the same amplitude for display purposes. Only for the DPSS controls the corresponding FFs are concentrated in both quadratures. 
\textbf{(h, i)} show the simultaneous experimental reconstruction of an engineered dephasing spectrum, $S_{z}(\omega)$, and an engineered control noise spectrum, $S_{\Omega}(\omega)$, that partly overlap in frequency space. Inset Bloch spheres show the effective rotation of a qubit state prepared in either $+x$ or $+y$ on the Bloch vector in three distinct frequency regions, which are separated by shaded regions indicating the spectral overlap between $S_{z}(\omega)$ and $S_{\Omega}(\omega)$. \textbf{(h)} Projective measurement data for each control waveform as a function of probe frequency. The error bars represent the variance over 100 individual time-domain noise realizations. Lines represent numerical simulations. \textbf{(i)} shows the reconstructed power spectra in both quadratures. The spectra and data have been normalized for display purposes, original spectrum amplitudes were $S_\Omega^{(\text{max})}(\omega) \approx \unit[450]{Hz}$ and $S_z^{(\text{max})}(\omega) \approx \unit[15]{Hz}$.  }
\label{fig:multi_axis_figure}
\end{figure*}

\subsection{Dephasing noise spectroscopy} 

\noindent
We now demonstrate the reconstruction of a noise spectrum resulting from an additive dephasing term $\beta_z(t)$ as in Eq. \eqref{eq:basic_hamiltonian}. Specifically, we engineer a complex, ``mixed'' spectrum that exhibits both a broadband $1/f$ component and discrete, narrowband spectral features. This spectrum is converted to a time-domain disturbance through an inverse Fourier transform and applied to the sensor via frequency modulation of the driving field, which is physically equivalent to an ambient dephasing field~\cite{Soare2014bath,BallClock}. Spectral reconstruction begins with application of appropriate finite-difference DPSS controls and execution of the above tomographic measurement protocol. 

Following the two-stage estimation procedure proposed in \cite{Norris2018}, we first perform a coarse sampling in frequency aimed to detect the presence of spectral structure, by using controls of duration {$\tau= N \Delta t = \unit[2.5]{ms}$ and bandwidth product $NW=4$, thereby achieving an effective sample bandwidth of \mbox{$f_B \equiv NW/\tau=\unit[~1.6]{kHz}$}.} If $\omega_s=2\pi f_s$ is the band-shift frequency, the resulting dephasing FFs are spectrally concentrated in a passband $B_s\equiv(\omega_s-2\pi f_B,\omega_s+2\pi f_B)$. Using the spectral concentration of the FFs to truncate the integral in \erf{ay} and assuming that $S_z(\omega)$ is locally flat in $B_s$, the dephasing spectrum is inferred from experimentally determined values of $\hat{\mathcal S}_y(\tau)$ using the relationship 
\begin{equation}
  \hat{S}_z(\omega_s)\approx \frac{\pi\hat{\mathcal S}_y(\tau)}{ \int_{B_s}d\omega\,F_z(\omega,\tau)}.
  \label{eq:spectrum_reconstruction}
\end{equation}
\noindent
We then supplement this initial coarse estimate of the spectrum with a fine scan, using enhanced spectral resolution in a region where prominent features deviating from a smooth trend are observed; this is achieved by adjusting the control duration to $\tau=\unit[5]{ms}$ and letting $NW=2$ to achieve $f_B=\unit[~0.4]{kHz}$ (Fig.~\ref{fig:dephasing_reconstruction}a). A Bayesian update (see Appendix \ref{app:bayes} for detail) is then used to combine the information from the coarse and fine scans to find the most likely spectral weight across the measurement range. Our experimental measurements and the associated reconstructions, shown in Fig. \ref{fig:dephasing_reconstruction}, provide both quantitative and qualitative agreement with the applied noise spectrum using no free parameters. We further provide a comparative analysis of the performance of our DPSS protocol against existing sensing protocols in Sec. \ref{cpmg_comparison}.

\subsection{Simultaneous multi-axis sensing} 

\noindent
We now address the challenge of multi-axis reconstruction of simultaneous, but statistically independent noise spectra, ${S}_{\Omega}(\omega)$ and ${S}_{z}(\omega)$; a scenario which, as we mentioned before, is commonly encountered for sensors in which the amplitude of the control itself suffers from fluctuations. Both spectra contribute to the evolution of the sensor's state, making direct spectral estimation from single-axis measurements difficult. Figure \ref{fig:multi_axis_figure}a schematically represents the measurement process including state preparation, controlled evolution, three-axis measurement and finally spectral reconstruction. Conventional control protocols struggle to accurately perform the final reconstruction because they are designed to be spectrally concentrated and tunable in at most one quadrature at a time (Fig.~\ref{fig:multi_axis_figure}b/e and c/f). In contrast, our finite-difference DPSS control, while being designed to ensure spectrally concentrated sensitivity to dephasing noise $\propto\sigma_{z}$, has the additional benefit that spectral concentration is also preserved for multiplicative control noise along $\propto\sigma_{x}$.  As a result, we see in Fig. \ref{fig:multi_axis_figure}b-g that, when compared to other common sensing protocols, \emph{only} the finite-difference DPSS modulation yields FFs which are simultaneously concentrated for both noise sources, thus enabling multi-axis spectral estimation (see also Sec. \ref{cpmg_comparison}). Specifically, this may be achieved by inferring the dephasing spectrum from $\hat{\mathcal{S}}_y(\omega)$, as previously described, and similarly inferring the amplitude spectrum by $$\hat{S}_\Omega(\omega_s)\approx \frac{\pi\hat{\mathcal S}_x(\tau) }{ \int_{B_s}d\omega\,F_\Omega(\omega,\tau)}.$$

To experimentally validate this approach, we engineer two Gaussian-shaped power spectral densities for both $S_{\Omega}(\omega)$ and $S_{z}(\omega)$, which partly overlap in frequency (shaded region, Fig.~\ref{fig:multi_axis_figure}i). Again, these spectra are converted to time-domain disturbances and applied to the sensor. We then perform tomographic measurements under application of these disturbances using finite-difference DPSS controls at $M=13$ different band-center frequencies (see Fig.~\ref{fig:multi_axis_figure}h). The amplitude of the finite-difference control waveform, $\Omega$, was selected so that $\int_0^\tau dt\,\Theta_\text{FD}(t)^2$ is constant for all $\omega_s$, ensuring that the peaks of the dephasing filters have constant magnitude. Since the amplitude FFs differ from the dephasing FFs by a factor of $\omega^2/4$, however, the peaks of the amplitude filters scale as $\omega_s^2$ and, thus, have varying magnitude depending on $\omega_s$.

Reconstructions based on the tomographic measurements obtained under simultaneous application of both noise spectra are shown in Fig.~\ref{fig:multi_axis_figure}i.  Data agree well with the applied noise spectra, including in the central frequency band of $6-12$ kHz, where both spectra contribute approximately equally to the overall sensor response. The larger error bars appearing for low frequency values arise due to uncertainty introduced by the fact that, as noted above, the amplitude-filter magnitude diminishes with reduced band-center frequency $\omega_{s}$, while the magnitude of the dephasing filters, on the other hand, is kept constant throughout the frequency scan range. Again, we stress that there are no free parameters used in representing the solid lines presented in Fig.~\ref{fig:multi_axis_figure}i.

\section{Assessment and comparison to 
\\ existing QNS approaches} 
\label{cpmg_comparison}

\begin{figure*}[t]
\includegraphics[width=1\textwidth]{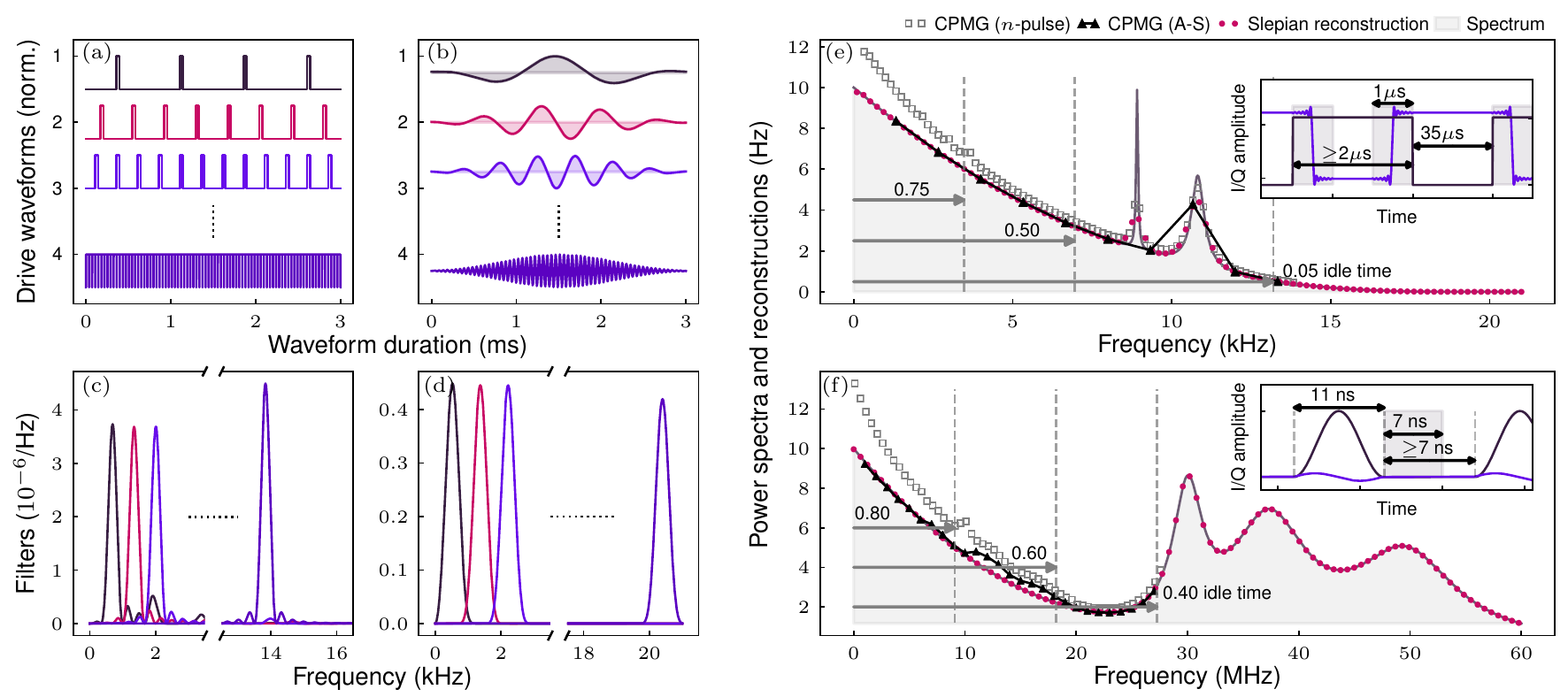}
 \caption{ Comparing DPSS and pulsed control for dephasing spectral estimation. In \textbf{(a, b)}, CPMG and finite-difference DPSS waveforms of total duration $\tau=\unit[3]{ms}$ are plotted in the time domain with the corresponding FFs shown in \textbf{(c, d)}.  The numerically simulated spectral reconstructions in \textbf{(e, f)} rely on three different QNS methods: DPSS (circles), CPMG-based $n$-pulse (squares) and A-S (triangles). In  \textbf{(e)}, these methods are numerically implemented with parameters relevant to our trapped ion platform. The $n$-pulse reconstruction uses CPMG sequences with pulse numbers ranging from $n=1$ to $n_\text{max}=83$. For the A-S, base sequences consisting of 2-pulse CPMG with duration $\tau_b/m$ for $m=1,\ldots, 10$ and $\tau_b = \unit[0.75]{ms}$ are repeated $M=4m$ times. The finite-difference DPSS with $k=0$, $N=600$ and $W=2/N$ are COS-shifted by $\omega_s/2\pi=0.1+0.211w$ kHz for $w=0,\ldots, 99$. All sequences have total duration $\tau= \unit[3]{ms}$ and the $\pi$-pulse duration is $\tau_\pi = \unit[35]{\upmu s}$ for the CPMG-based protocols. The inset shows a schematic $I/Q$ waveform that resembles the actual CPMG waveforms we use in experiments, with the shaded area highlighting the ringing response of the $I/Q$ baseband, which we suppress using microwave blanking markers. In \textbf{(f)}, the three QNS methods are implemented using the parameters of the superconducting qubit device in Ref. \cite{Sung2019}. The $n$-pulse CPMG sequences have pulse numbers ranging from $n=1$ to $n_\text{max}=550$, whereas the A-S method uses base sequences consisting of 2-pulse CPMG with duration $\tau_b/m$, repeated $M=10m$ times, with $\tau_b=\unit[1]{\upmu s}$ and $m=1,\ldots, 27$. For the finite-difference DPSS, $k=0$, $N=2000$, $W=2/N$ and the waveforms are COS-shifted by $\omega_s/2\pi=0.1+0.615w$ kHz for $w=0,\ldots, 99$. All sequences have total duration $\tau = \unit[10]{\upmu s}$. For the CPMG-based protocols, $\tau_\pi = \unit[11]{ns}$ and the buffer time is $\unit[7]{ns}$. The inset shows again a sample $I/Q$ waveform as used in \cite{Sung2019}. In both \textbf{(e)} and \textbf{(f)}, the dashed lines indicate the maximum achievable frequency $\omega_{\text{max}}$ relating to different ratios of idle vs. pulse time in the CPMG sequences (e.g., a value of $0.05$ means that $\unit[95]{\%}$ of the sequence consists of driven evolution). }
\label{fig::cpmg_comparison}
\end{figure*}

\noindent We have seen that DPSS control has, by design, superior spectral concentration compared to control sequences more commonly used in QNS, such as the CPMG sequence shown in Fig. \ref{fig:finite_diff_filters} \cite{Meiboom1958}, and that spectrally concentrated filters can be successfully deployed in experiment. Here, we present a more detailed comparison between the reconstruction capabilities of our DPSS method versus more established QNS protocols, in a setting where only additive dephasing noise is present.  As we already emphasized, our DPSS protocol is unique in its ability to simultaneously permit reconstruction of multiplicative noise in the control quadrature. 

Specifically, we focus on two pulsed QNS protocols: dynamical-decoupling QNS, first employed by Bylander \emph{et al} \cite{Bylander2011}, which estimates the dephasing noise spectrum $S_z(\omega)$ using $n$-pulse decoupling sequences (CPMG henceforth), and the frequency-comb approach developed by Alvarez and Suter \cite{Alvarez_Spectroscopy} (``A-S'' in what follows), which employs sequence repetition to ensure the emergence of a comb in frequency space, with narrow teeth probing $S_z(\omega)$ at the corresponding ``harmonic'' frequencies.  First, we examine the limitations of these protocols in terms of \emph{scan range} (the frequency range over which $S_z(\omega)$ can be reconstructed) and the \emph{sampling resolution} (the frequency separation between points in the reconstruction). To demonstrate the practical implications of spectral concentration, scan range and sampling resolution, we then simulate CPMG- and DPSS-based QNS using realistic experimental parameters taken from two distinct platforms: our ion-trap setting and the superconducting qubit device from Ref. \cite{Sung2019}.  Overall, subject to comparable design constraints we find that the key advantages of DPSS protocols are:
\begin{enumerate}
    \item     
    Reduced spectral leakage\vspace*{-1mm};        
%   \item 
%   Ability to reconstruct (multiplicative) amplitude noise in addition to dephasing noise; \vspace*{-1mm}and     
%% LV: This is true but not part of the comparison, I'd say    
    \item 
    Increased scan range and improved spectral resolution through arbitrary waveform control.
\end{enumerate}
%The first point arises directly from the nature of the DPSS waveforms, as we discussed in detail in previous sections.
%, while the last point illustrates a very common experimental subtlety, as we elucidate next. 
We illustrate our comparison and the above claims next.

Consider first CPMG-based QNS.
%The first of the CPMG-based QNS methods we consider is the $n$-pulse CPMG QNS.
%approach first employed by Bylander \emph{et al}, which estimates the dephasing noise spectrum using $n$-pulse CPMG control sequences \cite{Bylander2011}. 
In this case, each $n$-pulse CPMG sequence has a fixed duration $\tau$ and consists of $n$ $\pi$-pulses applied about $\sigma_x$, separated by a $n$-dependent inter-pulse spacing (the duration from the center of a pulse to the center of a neighboring pulse), which we call $\Delta \tau_n$. Representative sequences for rectangular pulses and varying $n$ are depicted in Fig. \ref{fig::cpmg_comparison}a.  For these sequences, the FF of interest is $F_{zz}(\omega,\tau)\equiv\int_0^\tau\!\!\mathrm{d}s\,e^{i\omega s}\cos\Theta(s)$ which, using Eq. \eqref{eq:error_vector_components_time_z}, enters the qubit dynamics through 
\begin{align}
\mathcal{S}_z(\tau)
=\,\frac{1}{\pi}\int_{0}^\infty\!\!\mathrm{d}\omega\, |F_{zz}(\omega,\tau)|^2S_z(\omega).
\label{eq::az2}
\end{align}
Let  $\tau_\pi$ denote the $\pi$-time, that is, the time required to apply a $\pi$-pulse. In the case where the $\pi$-pulses are (nearly) instantaneous, meaning that $\Delta \tau_n\approx \tau/n\gg \tau_\pi$, the rotation angle $\Theta(t)$ is always an integer multiple of $\pi$, implying that $\cos\Theta(t)\rightarrow y(t)$, where the switching function $y(t)\in\{\pm 1\}$ changes sign each time a $\pi$-pulse is applied. For each $n$, this creates an approximate bandpass FF peaked at $ \omega_p\approx n\pi/\tau$. The dephasing noise spectrum is then estimated using
\begin{equation}
  \hat{S}_z(n\pi/\tau)\approx \frac{\pi\hat{\mathcal S}_z(\tau)}{ \int_{B_n}d\omega\,F_z(\omega,\tau)},
\end{equation}
where $B_n$ is the passband containing the main peak of the $n$-pulse CPMG filter, as depicted in Fig. \ref{fig::cpmg_comparison}c.
Since the estimate is based on the spectral weight of the FF within the passband, spectral weight outside the passband constitutes leakage.

In any experimental implementation, $\pi$-pulses are, of course, never instantaneous. As we shall show, a nonzero $\tau_\pi$ constrains the reconstruction capabilities of the $n$-pulse method. Assuming square $\pi$-pulses of amplitude $\Omega_\pi\equiv\pi/\tau_\pi$, we find that $\cos\Theta(t)= Y(t)H(\tau-t)$, where $H(t)$ is the Heaviside step function and $Y(t)$ is periodic in $2\tau/n$,
%\begin{align*}
%Y(t)=\begin{cases}
%  1 & t\in[2k\tau/n,2k\tau/n+\Delta \tau_n/2),\\
%  \cos [\Omega_\pi (t - t'_1)]& t\in [2k\tau/n+\Delta \tau_n/2, \\
%  & \qquad 2k\tau/n+\Delta \tau_n/2+\tau_\pi),\\
%  -1& t\in[2k\tau/n+\Delta \tau_n/2+\tau_\pi,\\
%  & \qquad 2k\tau/n+3\Delta \tau_n/2+\tau_\pi),\\
%  -\cos [\Omega_\pi (t - t'_2)]& t\in[2k\tau/n+3\Delta \tau_n/2+\tau_\pi,\\
%  & \qquad 2k\tau/n+3\Delta \tau_n/2+2\tau_\pi)\\
%  1& t\in[2k\tau/n+3\Delta \tau_n/2+2\tau_\pi,\\
%  & \qquad 2(k+1)\tau/n),
% \end{cases}
%  \end{align*} 
 \begin{align*}
  Y(t)\!=\!\begin{cases}
  +1 & \!t\in[2k\tau/n, t_{1,k} ) , \\
% 2k\tau/n+\Delta \tau_n/2),\\
  + \cos [\Omega_\pi (t - t_{1,k})]&\! t\in [t_{1,k} , t_{1,k}+\tau_\pi), \\
%  2k\tau/n+\Delta \tau_n/2, \\
%  & \qquad 2k\tau/n+\Delta \tau_n/2+\tau_\pi),\\
  -1& \!t\in[ t_{1,k}+\tau_\pi, t_{2,k}), \\
%  2k\tau/n+\Delta \tau_n/2+\tau_\pi,\\
%  & \qquad 2k\tau/n+3\Delta \tau_n/2+\tau_\pi),\\
  -\cos [\Omega_\pi (t - t_{2,k})]& \! t\in[t_{2,k}, t_{2,k}+\tau_\pi), \\
%  2k\tau/n+3\Delta \tau_n/2+\tau_\pi,\\
%  & \qquad 2k\tau/n+3\Delta \tau_n/2+2\tau_\pi)\\
  1& \! t\in[t_{2,k} +\tau_\pi, 2(k+1)\tau/n), 
  %2k\tau/n+3\Delta \tau_n/2+2\tau_\pi,\\
%  & \qquad 2(k+1)\tau/n),
\end{cases}
\end{align*} 
where $k\in {\mathbb N}$  and we have introduced the shorthands $t_{1,k} \equiv 2k\tau/n+\Delta \tau_n/2$ and 
%$t_{2,k} \equiv 2k\tau/n+3\Delta \tau_n/2+\tau_\pi$. 
$t_{2,k} \equiv t_{1,k}+ \Delta \tau_n +\tau_\pi$. In terms of the Fourier coefficients $a_\nu\equiv\frac{n}{\tau}\int_0^{2\tau/n}\mathrm{d}s\,Y(s)\cos(\pi n \nu s/\tau)$, we can then write the $n$-pulse CPMG FF as \cite{GreenNJP2013, Szankowski2017}
\begin{align*}
|F_{zz}(\omega,\tau)|^2=&\bigg|\,\pi\tau\sum_{\nu=1}^\infty a_\nu \Big[ e^{-i(\omega-\pi n\nu/\tau)}\text{sinc}\Big(\frac{\omega\tau-\pi n\nu}{4\pi}\Big) \\
&+e^{-i(\omega+\pi n\nu/\tau)}\text{sinc}\Big(\frac{\omega\tau+\pi n\nu}{4\pi}\Big)\Big]\,\bigg|^2,
\end{align*}
which consists of a series of peaks centered at $\omega_\nu\equiv \pi n\nu/\tau$, with magnitudes dictated by $a_\nu$. 

In this expression for the filter function, the largest Fourier coefficient is $a_1$, corresponding to the projection of $Y(t)$ onto $\cos(\pi n t/\tau)$, which is also periodic in $2\tau/n$. As a consequence, $|F_{zz}(\omega,\tau)|^2$ has a maximum peak centered at $\omega_1=n\pi/\tau=\omega_p$, like the instantaneous case.
The largest possible scan range is then $\omega_\text{max}=n_\text{max}\pi/\tau=\pi/\Delta\tau_\text{min}$, where the maximum number of pulses $n_\text{max}=\tau/\Delta\tau_\text{min}$ is set by the minimum interpulse spacing, $\Delta\tau_\text{min}$. Due to the finite widths of the $\pi$-pulses, $\Delta\tau_\text{min}>\tau_\pi$, meaning that the scan range is ultimately limited by the $\pi$-time. In practice, other experimental non-idealities, such as transients in the drive response, often extend $\Delta\tau_\text{min}$ beyond $\tau_\pi$, so that it never actually achieves this lower bound. From the locations of the peaks, we can also determine the sampling resolution, 
%for the $n$-pulse method, 
which is $\omega_\text{res}=\pi/\tau$.

%In addition to the $n$-pulse method, we also consider the frequency-comb approach developed by Alvarez and Suter \cite{Alvarez_Spectroscopy}, referred to here as ``A-S''. 
In QNS via the A-S method, a fixed base sequence of duration $\tau_b$ is repeated $M \gg 1 $ times, which generates a series of comb-like peaks in $|F_{zz}(\omega,M\tau_b)|^2$ that are centered at the harmonic frequencies or integer multiples of $\omega_0\equiv 2\pi/\tau_b$. In \erf{eq::az2}, restricting the bounds of integration to a finite range and approximating the peaks in $|F_{zz}(\omega,M\tau_b)|^2$ as delta functions produces then a linear equation,
\begin{align}
\label{eq::ASLinEq}
\mathcal{S}_z(M\tau_b)=\frac{2M}{\tau_b}\sum_{m=1}^{m_\text{max}}|F_{zz}(m\omega_0,\tau_b)|^2S_z(m\omega_0),
 \end{align} 
which couples the FF corresponding to a single repetition to the spectrum evaluated at a finite set of harmonic frequencies. Repeating this procedure for at least $m_\text{max}$ different sequences with base sequence durations $\tau_b$ or some fraction of $\tau_b$, i.e., $\tau_b/m$ for integer $m>0$, creates a system of linear equations that can be inverted to estimate $\mathcal{S}_z(\omega)$ at $\omega \in \{\omega_0,2\omega_0,\ldots,m_\text{max}\omega_0\}$.  Because the inversion procedure takes into account the signal picked up by the higher-frequency peaks of each FF, the A-S method is more robust to spectral leakage than the $n$-pulse CPMG is.

Since the spectrum is estimated at the harmonic frequencies set by $\tau_b$, the sampling resolution of the A-S method is naturally given by $\omega_\text{res}=\omega_0\equiv 2\pi/\tau_b$. The scan range, $\omega_\text{max}=m_\text{max}\omega_0$, depends on the total number of sequences used in the reconstruction. Determining the upper bound for $m_\text{max}$, however, is a more subtle issue and generally depends on the particular set of sequences employed (see also Sec. V. B in Ref. \cite{NorrisNG} for an expanded discussion, applicable to more general frequency-comb-based protocols employing a set of different base sequences).  Because they ensure a well-conditioned inversion, we consider the sequences that A-S originally used in Ref. \cite{Alvarez_Spectroscopy}, which consist of repeated applications of 2-pulse CPMG with variable base duration $\tau_b/m$, for $m=1,\ldots,m_\text{max}$. The maximum number sequences is then constrained by the minimum spacing between the two pulses in the CPMG base sequences, $m_\text{max}=\tau_b/2\Delta\tau_\text{min}<\tau_b/2\tau_\pi$, with an upper bound set by the $\pi$-time. Consequently, the achievable scan range is given by $\omega_\text{max}=\pi/\Delta\tau_\text{min}<\pi/\tau_\pi$, which is equivalent to the $n$-pulse method.

A key advantage of our DPSS approach is that the \emph{scan range and sampling resolution can be tuned independently}.  For both of the above pulsed protocols, by contrast, we see that the scan range is restricted by the minimum interpulse spacing.  The minimum interpulse spacing ultimately influences the total sequence duration for the $n$-pulse method and the base-sequence duration for A-S, which in turn constrain the sampling resolution. In the DPSS case, the scan range and sampling resolution are determined by the time increment $\Delta t$ and the shift frequencies $\omega_s$, respectively, each of which can be adjusted separately in a DPSS waveform. The Nyquist frequency, $\omega_N\equiv \pi/\Delta t$, sets the scan range for the DPSS. Beyond $\omega_\text{max}=\omega_N$, the spectrum cannot be estimated due to aliasing effects in the FFs \cite{Norris2018}. In practice, the smallest possible time increment is given by $\Delta t_\text{min}=2\pi/\omega_\text{SR}$, where $\omega_\text{SR}$ is the maximum sampling rate of the arbitrary waveform generator (AWG). Consequently, the maximum achievable scan range is $\omega_\text{max}=\pi/\Delta t_\text{min}=\omega_\text{SR}/2$. For the DPSS, the sampling resolution is set by the separation between neighboring shift frequencies $\omega_\text{res}=|\omega_{s,1}-\omega_{s,2}|$.

Figure \ref{fig::cpmg_comparison} e) and f) show numerically simulated spectral reconstructions using the $n$-pulse CPMG, A-S, and DPSS methods with parameters representative of the ion-trap and the superconducting qubit platforms, respectively.  The waveforms in each reconstruction all have the \emph{same duration}, $\tau = \unit[3]{ms}$ in e) and $\tau = \unit[10]{\upmu s}$ in f). In a realistic implementation, this would ensure that the qubit is subject to the same amount of noise during each shot of the experiment for each of the QNS methods. In both reconstructions, spectral leakage of the $n$-pulse CPMG FFs causes this approach to overestimate the actual noise strength at lower frequencies. The primary source of error for the A-S method stems from ignoring higher-order harmonics to generate the linear equation in \erf{eq::ASLinEq}. Because the sequence of harmonics is truncated at $\omega_\text{max}=m_\text{max}\omega_0$, the inversion procedure does not account for spectral weight of $S_z(\omega)$ at $\omega > \omega_\text{max}$, biasing the reconstruction. This effect is not apparent in \ref{fig::cpmg_comparison} e), since the spectral weight is minimal beyond $\omega_\text{max}/2\pi\approx 13$ kHz. The spectrum in  \ref{fig::cpmg_comparison} f), on the other hand, has considerable spectral weight beyond $\omega_\text{max}/2\pi\approx 27$ MHz, which introduces error in the 10-20 MHz range. The DPSS are relatively free of leakage and truncation error, resulting in accurate spectral reconstructions in both e) and f).

In addition to leakage and truncation error, the CPMG reconstructions at fixed sequence duration are limited in terms of scan range and sampling resolution by  intrinsic  features of the pulse implementation, as we have discussed.
In the case of the ion-trap-inspired simulations (Fig. \ref{fig::cpmg_comparison}e), pulses are implemented using rectangular $I/Q$ waveform shapes with $\tau_\pi \approx \unit[35]{\upmu s}$. However, to suppress the ringing response of the $I/Q$ modulator to sharp waveform edges, additional blanking samples must be added to the waveforms, which adds $1 {\upmu s}$ of buffer time to either side of the $\pi$-pulse and produces a minimum interpulse spacing of $\Delta\tau_\text{min} \approx \unit[37]{\upmu s}$. The maximum scan range of both the $n$-pulse and A-S methods is, thus, $\omega_\text{max}/2\pi=1/2\Delta\tau_\text{min}\approx \unit[13]{kHz}$.
For the DPSS reconstruction in Fig. \ref{fig::cpmg_comparison} e), $\Delta t=\tau/N= \unit[5]{\upmu s}$, which gives a scan range $\omega_\text{max}/2\pi=100$ kHz.  In terms of sampling resolution, the DPSS and $n$-pulse CPMG reconstructions are comparable. For the A-S protocol, however, the need to apply repetitions of the 2-pulse CPMG base sequences (at least $M=4$ repetitions, in our example) constrains the maximum base sequence duration to $\tau_b=\tau/4=\unit[0.75]{ms}$.  The resulting sampling resolution of $\omega_\text{res}/2\pi\approx 1.3$ kHz is insufficient to resolve the first peak of the spectrum, unlike the DPSS and $n$-pulse CPMG. For the superconducting-qubit-inspired parameters considered in panel f), a minimum buffer time of $ \unit[7]{ns}$ is required to ensure full separation between consecutive $\pi$-pulses of width $\tau_\pi=\unit[11]{ns}$  \cite{Sung2019}. For the $n$-pulse and A-S protocols, this produces a minimum interpulse spacing of $\Delta\tau_\text{min}=\unit[18]{ns}$, limiting the scan range to $\omega_\text{max}/2\pi \approx \unit[27.7]{MHz}$. The scan range of the DPSS reconstruction, in contrast, is $\omega_\text{max}/2\pi=100$ MHz, corresponding to a time increment of $\Delta t=\unit[5]{ns}$.  The sampling resolutions of the three protocols are comparable for the superconducting qubit parameters. In neither example did we reach the maximum possible scan range for the DPSS set by the AWG, which is $\omega_\text{max}/2\pi=5$ MHz for the trapped ion and $\omega_\text{max}/2\pi=500$ MHz for the superconducing qubit device.

The numerical reconstructions in Fig. \ref{fig::cpmg_comparison} employed sequences for the CPMG-based protocols in which the pulses were separated $\Delta \tau_\text{min}$. In other words, these simulations went to the limit of what is achievable using the pulse parameters derived from the trapped ion and superconducting qubit platforms. Often, in an actual experimental implementation, it is desirable to keep the separation between pulses above the minimum interpulse spacing by imposing some fraction of free evolution, or \emph{idle} time, in the control sequence. If $n$ is the total number of pulses in a sequence, the fraction of idle time is given by $r_\text{idle} \equiv 1 - n \Delta \tau_\text{min} / \tau$, provided $n\leq\tau/\Delta \tau_\text{min}$. If the sequence is entirely free evolution, for example, $r_\text{idle}=1$. Imposing some fraction of idle time necessarily increases the spacing between pulses and limits the scan range to $\omega_\text{max}= \pi(1 - r_\text{idle})/\Delta \tau_\text{min}$. To show how imposing idle time would affect spectral reconstruction in the trapped ion and superconducting qubit systems, Fig. \ref{fig::cpmg_comparison} e) and f) shows scan ranges corresponding to different values of $r_\text{idle}$  in the limit where $\Delta \tau_\text{min}=\tau_\pi$. It is immediately apparent that adding even a modest amount of idle time in CPMG-based spectroscopy can significantly reduce the accessible scan range of the reconstruction.

In closing our comparison, we note that in addition to pulsed QNS protocols as we focused on, 
%Although we have focused our comparison on pulsed CPMG-based protocols for reconstructing the dephasing noise spectrum, 
other QNS approaches exist for dephasing noise, such as the spin-locking method used in Refs. \cite{Yan2013,Yan2018} and recently extended to two qubits  \cite{Uwe}. This method has similarities to our finite-difference DPSS approach in that it characterizes dephasing noise along $\sigma_z$ by continuous driving along an orthogonal spin component, $\sigma_x$, for instance. The waveform used in spin-locking is of constant amplitude, i.e., $H_{\text{ctrl}} (t) =\Omega\sigma_x/2$, which produces a dephasing FF equivalent to that of free evolution, with the main peak translated by $\pm\Omega$ in the frequency domain. The FF produced by free evolution and other ``flattop'' waveforms do not offer significant 
%optimally 
%% LV: only DPSS is optimal! :)
leakage suppression \cite{Frey2017}. Similar to pulsed control, an added benefit of the finite-difference DPSS approach is the ability to additionally characterize 
%dephasing and 
multiplicative amplitude noise, which is not possible with conventional spin-locking techniques.

\section{Outlook} 

\noindent
In summary, we demonstrated the implementation of a continuously driven control protocol for quantum sensing which exhibits optimal spectral concentration in the dephasing quadrature $\propto \sigma_z$, for controls driven in $\sigma_x$ or $\sigma_y$, as is the typical setting in many sensing applications. The continuous nature of these controls offers superior flexibility compared to standard pulsed control and the resolution of the reconstruction is arbitrarily tunable (subject to sampling resolution of the arbitrary waveform generators). Additionally, our controls exhibit spectral concentration in both the dephasing and control quadrature simultaneously and can therefore be employed to produce a minimally biased sensor in applications that suffer from time-dependent control noise which distorts the target dephasing signal. This allows for simultaneous, multi-axis spectral estimation of both the control noise and the dephasing noise power spectral densities. 

Future work will see the extension of our protocol to different Hamiltonian models for describing the interaction between control and noise -- in particular, the case of control noise which couples additively, as encountered with cross-talk. Furthermore, we are interested in different models for the noise that relax assumptions about the noise process, to include e.g., non-classical \cite{Yan2018} or non-Gaussian noise \cite{NorrisNG,Sung2019} in both the control and dephasing quadrature. Finally, we believe designing controls and filters for multi-qubit operations to sense noise and unwanted cross-talk across an array of qubits is an important direction to pursue in order to perform spatial as well as temporal noise reconstruction, both of which would leverage the spectrally concentrated properties of the DPSS control sequences.

%\vfill
 
\section*{Acknowledgements} 
It is a pleasure to thank D. Lucarelli for pointing us towards the DPSS and for helpful discussions, and S. Mavadia for early contributions, in particular to the idea of pulsed DPSS protocols. We also acknowledge J. Wakulicz and A. Singh for assisting in the experimental data acquisition process. This work was partially supported by the ARC Centre of Excellence for Engineered Quantum Systems CE110001013, the US Army Research Office under Contract W911NF-12-R-0012, IARPA via Department of Interior National Business Center contract number 2012-12050800010, and a private grant from H. \& A. Harley.

\iffalse
\medskip
\noindent 
{\bf Author contributions.} 
V. F. developed experimental hardware, built the experimental control system, and obtained the presented data with theoretical techniques developed by L.M.N. and L.V.. Numerical simulations and data analysis were performed by V. F., L.M.N., and L.V. M.J.B. conceived the experiment and led development of the experimental system. V. F., L.M.N., L.V., and M.J.B. jointly wrote the manuscript.

\medskip

\noindent 
{\bf Additional information.} Experimental and numerically simulated data, as well as simulation code are available from the authors upon request.
\fi

\appendix

\begin{widetext}

\section{Finite-difference control with embedded dynamical decoupling}
\label{app::embed}

\noindent 
As discussed in the main text, when truncating the Magnus expansion to the first order is viable, we can obtain the signal projection via projective measurements along three-axes, via $\mathcal{S}_y(\tau)\approx[1+P(\uparrow_y,\tau)-P(\uparrow_x,\tau)-P(\uparrow_z,\tau)]/2$. Beyond the weak noise limit, this procedure is complicated by the presence of higher-order terms in the Magnus expansion. Specifically, 
\begin{align}
\frac{1+P(\uparrow_y,\tau)-P(\uparrow_x,\tau)-P(\uparrow_z,\tau)}{2}\approx\mathcal{S}_y(\tau)+2\expect{a_y^{(1)}(\tau)a_y^{(3)}(\tau)}
-\frac{1}{3}\expect{a_y^{(1)}(\tau)^2a_z^{(1)}(\tau)^2},\label{eq::SyHigherOrder}
\end{align}
where we have shown terms up to order $\tau^4$. Fortunately, the higher-order terms above depend on functionals of the form $\int_0^\tau ds\,(\cdot)\,\cos\Theta(s)$, which are absent in $\mathcal{S}_y(\tau)$, as seen in \erf{ay}.  Observe that under the transformation $\Theta(t)\mapsto\pi-\Theta(t)$, $\cos\Theta(t)\mapsto-\cos\Theta(t)$ whereas $\sin\Theta(t)\mapsto\sin\Theta(t)$. As we show below, this allows for the possibility to suppress the higher-order terms through dynamical decoupling targeted at $\cos\Theta(t)$, while preserving $\mathcal{S}_y(\tau)$.

For $N$ such that $N/4$ is an integer, consider the following modified version of the finite-difference waveform in \erf{eq::SuppFD}:
\begin{align*}
\Omega(t)=\begin{cases}
V_0'\equiv V_0, & t\in\Delta t\,[0,1),\\
V_1'\equiv V_1\!-\!V_0, & t\in\Delta t\,[1, 2),\\
\;\;\;\;\;\;\;\;\vdots&\;\;\;\;\;\;\vdots\\
V_{N/4-1}'\equiv V_{N/4-1}\!-\!V_{N/4-2}, & t\in\Delta t\,[N/4\!-\!2, N/4\!-\!1),\\
V_{N/4}'\equiv \pi/\Delta t-V_{N/4}\!-\!V_{N/4-1}, & t\in\Delta t\,[N/4\!-\!1,N/4),\\
V_{N/4+1}'\equiv V_{N/4}\!-\!V_{N/4+1}, & t\in\Delta t\,[N/4, N/4\!+\!1),\\
\;\;\;\;\;\;\;\;\vdots&\;\;\;\;\;\;\vdots\\
V_{3N/4-1}'\equiv V_{3N/4-2}-V_{3N/4-1}, & t\in\Delta t\,[3N/4\!-\!2, 3N/4\!-\!1),\\
V_{3N/4-1}'\equiv V_{3N/4}+V_{3N/4-1}-\pi/\Delta t, & t\in\Delta t\,[3N/4\!-\!1, 3N/4),\\
V_{3N/4}'\equiv V_{3N/4}-V_{3N/4+1}, & t\in\Delta t\,[3N/4, 3N/4\!+\!1),\\
\;\;\;\;\;\;\;\;\vdots&\;\;\;\;\;\;\vdots\\
V_{N}'\equiv V_{N}-V_{N-1}, & t\in\Delta t\,[N, N\!-\!1).
\end{cases}
\end{align*}
If $\Omega(t)$ is such that $\Theta(t)\ll\pi/2$ for all $t$, this waveform produces 
\begin{align*}
\sin\Theta(m\Delta t)\approx V_{m-1}\Delta t, \qquad \cos\Theta(m\Delta t)\approx\begin{cases}
1,&t\in\Delta t[0,N/4),\\
-1,&t\in\Delta t[N/4,3N/4),\\
1,&t\in\Delta t[3N/4,N).\end{cases}
\end{align*}
The first expression is characteristic of ordinary finite-difference modulation, while the second approximates the switching function of a CPMG sequence. Thus, $\mathcal{S}_y(\tau)$ takes the form of \erf{ay}, where $F_z(\omega,\tau)$ is a spectrally concentrated finite-difference filter, and the higher order terms in \erf{eq::SyHigherOrder} are suppressed. Using a similar procedure, we can generate $\cos\Theta(t)$ with sign changes at some set of arbitrary times $\{m_1\Delta t,\cdots ,m_n \Delta t\}$, allowing for the possibility of higher-order decoupling sequences such as concatenated decoupling. When the higher-order terms in \erf{eq::SyHigherOrder} are sufficiently suppressed, the signal projection can be obtained from the usual expression, $\mathcal{S}_y(\tau)\approx[1+P(\uparrow_y,\tau)-P(\uparrow_x,\tau)-P(\uparrow_z,\tau)]/2$, to good approximation. In doing so, care should be taken to ensure that the presence of periodicities in the applied pulses control does not generate harmonic components which could re-introduce appreciable spectral leakage in the target frequency range of reconstruction.

\section{Experimental platform}
\label{app:exp_platform}

\noindent Our experimental testbed consists of a single trapped $^{171}$Yb$^+$ ion in a linear Paul trap with qubit transition realized through the hyperfine splitting of the $S_{1/2}$ ground state. State initialization and readout is performed optically using a $\unit[369]{nm}$ laser with $\unit[935]{nm}$ and $\unit[638]{nm}$ repump lasers. Typical readout fidelities are around $\unit[99.7]{\%}$. More information about the optical and trap setup can be found in previous works \cite{MavadiaNatComms2017, SoareNatPhys2014}.

The qubit transition frequency is at $\unit[\sim 12.6]{GHz}$ which we drive using the amplified output of the commercial Keysight E8267D vector signal generator (VSG). A waveguide-to-coax converter creates free space microwaves that are routed through one of the trap viewports to the ion. Typical $\pi$-times are about $\unit[30-40]{\upmu s}$ with $T_2$ times of $\unit[200]{ms}$. The VSG allows for programmable, digital $I/Q$ modulation of the carrier frequency, enabling arbitrary control in both the $x$- and $y$-axis of the Bloch sphere through the effective Hamiltonian $H_{\mathrm{ctrl}} (t) = \Omega(t) [\cos \varphi(t) \sigma_x + \sin \varphi(t) \sigma_y]$. The driving amplitude, $\Omega(t)$, is set by the magnitude of the $I$ and $Q$ waveforms via $\Omega(t) = \sqrt{I^2(t) + Q^2(t)}$, and the angle between $I$ and $Q$ determines the phase $\varphi(t) = \tan[Q(t)/I(t)]$. It is our convention to let the $I$ quadrature correspond to the $x$-quadrature (and equivalently the $Q$ quadrature to $y$), hence when $Q(t)=0$ we say we are driving an $x$-rotation. The DPSS waveforms are calculated numerically on the experiment PC and then uploaded to the VSGs internal $I/Q$ DACs. All waveforms in this work are symmetric about zero and implement an identity operation. Separate waveforms to implement fast $\pi/2$ rotations before and after the DPSS waveform to perform the three-axis measurement are concatenated with the DPSS waveforms on the VSG, yielding three $I/Q$ waveform sequences for each DPSS. 

For experiments with engineered noise, amplitude noise is generated digitally and added to the DPSS waveforms before the upload. Dephasing noise is added using a Keysight 33600A AWG that likewise produces an analog waveform from a digital input, which is then fed to the external frequency modulation (FM) input of the VSG. This effectively implements a dephasing noise term $\beta_z(t)\sigma_z$, where $\beta_z(t)$ is the waveform produced by the AWG. For the experiments mapping out FFs, as reported in Fig. 1 in the main text, $\beta_z(t)$ was implemented as a single-frequency sine-wave with variable phase $\phi$, that was linearly sampled from 0 to $2\pi$. Each point in the reconstruction consists of an average over five individual measurements taken with different values of $\phi$, so that $\langle \beta_z (t)\beta_z(t')\rangle \propto \cos (\omega_{\rm{sid}} (t-t'))$. For all other experiments that used engineered dephasing noise with a target spectrum $S_z^{\rm{t}}(\omega)$, we used a waveform $\beta_z(t) \propto \sum_i \sqrt{S_z^{\rm{t}} (\omega_i)} \cos(\omega_i t +\phi_i)$. Averaging over phases in the time-domain results in the frequency-domain spectrum $S_z(\omega) \propto \sum_i S^{\rm{t}}_z(\omega_i) \delta(\omega-\omega_i)$. We call each $\beta_z(t)$ that is calculated with a fixed set of random phases $\{\phi_i\}$ a single ``noise realization''.

\section{Bayesian spectral reconstruction procedure}
\label{app:bayes}

\noindent The two-step Bayesian spectral reconstruction depicted in Fig. 2 of the main text is based on a procedure for the detection of peaks and narrowband spectral features detailed in Ref. \cite{Norris2018}.  This procedure involves an initial detection stage, in which wide-band DPSS FFs are used for a coarse reconstruction of the target noise spectrum. Statistically significant peaks or bumps in the coarse reconstruction signify the presence of narrowband spectral features, which are resolved with a subsequent fine-sampling of the spectrum using narrowband DPSS FFs. Measurements from the initial detection state determine a prior estimate of the spectrum, which is updated based on the subsequent measurements using narrowband Slepian filters.

In the present experiment, finite-difference modulation was used to generate 9 wide-band (or ``coarse'', $c$) $k=0$ DPSS FFs, centered at $f_s =0, 2.1, 4.2, 6.2,  8.3, 10.4,$ $ 12.5, 14.6, 16.7$ kHz   [Fig.  2(a) of the main text]. We denote the wide-band FF centered at $\omega_s=2\pi f_s$ by $F_z^{c,s}(\omega,\tau)$.
For each $\omega_s$, the signal projection is related to the dephasing spectrum and $F_z^{c,s}(\omega,\tau)$  by
\begin{align}
\label{eq::ay2}
\mathcal{S}_y^{c,s}(\tau)=\,\frac{1}{\pi}\int_{0}^\infty\!\!\!\!\mathrm{d}\omega\,S_z(\omega)F_z^{c,s}(\omega,\tau)
\approx \,\frac{1}{\pi}\int_{B_s}\mathrm{d}\omega\,S_z(\omega)F_z^{c,s}(\omega,\tau),
\end{align}
where we have used the fact that each FF is spectrally concentrated in a band $B_s\equiv(\text{max}\{0,\omega_s-\Delta b\},\,\omega_s+\Delta b)$ and $\Delta b\equiv 2\pi W/\Delta t$ to restrict the domain of integration in the second line. Using this expression, the spectrum at $\omega_s$ can be estimated by $\hat{S}(\omega_s)\equiv\hat{\mathcal{S}}_y^{c,s}(\tau)/A_{c,s}$, where $\hat{\mathcal{S}}_y^{c,s}(\tau)$ is the measured value of $\mathcal{S}_y^{c,s}(\tau)$ and $A_{c,s}\equiv\frac{1}{\pi}\int_{B_s}d\omega\,F_z^{c,s}(\omega,\tau)$. The estimate $\hat{S}(\omega_s)$, which is known as the $k=0$ eigenestimate of the spectrum at $\omega_s$, is in good agreement with the actual spectrum provided that  $S_z(\omega)$ does not vary appreciably within $B_s$ \cite{thompson_multitaper,Norris2018}.  The eigenestimates for each $\omega_s$, plotted in Fig. 2 (b) of the main text, closely match the actual spectrum at all frequencies except for the $8-12$ kHz region, in which the spectrum has narrowband features and, consequently, varies significantly within $B_s$.

The two-step Bayesian procedure returns estimates of the spectrum in each of the 19 ``segments'' depicted in Fig. \ref{fig::segments}. The segments, which we denote by $\{\sigma_\ell\,|\ell=1,\ldots,19\}$, are narrower in the $8-12$ kHz region in order to resolve the spectral features absent in the initial eigenestimates. 
First, we determine the most probable estimate of the spectrum in the 19 segments based on the 9 original measurements of $\hat{\mathcal{S}}_y^{c,s}(\tau)$ using the wide-band FFs. This will serve as a prior, which will be updated based on subsequent measurements. To establish the prior, we first
discretize \erf{eq::ay2}, which yields
\begin{align*}
\hat{\mathcal{S}}_y^{c,s}(\tau)\approx \sum_{\ell=1}^{19}\frac{1}{\pi}\int_{\sigma_\ell}\!\!\!\!\mathrm{d}\omega\,S_z(\omega)F_z^{c,s}(\omega,\tau)
\approx \sum_{\ell=1}^{19}S_\ell\,\frac{1}{\pi}\int_{\sigma_\ell}\!\!\!\!\mathrm{d}\omega\,F_z^{c,s}(\omega,\tau),
\end{align*}
where $S_\ell$ is the average value of $S_z(\omega)$ in segment $\sigma_\ell$. If we gather the values of $\hat{\mathcal{S}}_y^{c,s}(\tau)$ for each wide-band filter into a $9\times 1$ vector, $\vec{\mathcal{S}}_y^c$, we can cast the expression above into a matrix equation $\vec{\mathcal{S}}_y^c=\mathbf{F}^c\vec{S},$
where $\vec{S}=(S_1,\ldots,S_{19})^T$ and $\mathbf{F}^c$ is a $9\times 19$ ``filter matrix'' with elements depending on the wide-band FFs, $(\mathbf{F}^c)_{s,\ell}=\frac{1}{\pi}\int_{\sigma_\ell}\!\!\!\!\mathrm{d}\omega\,F_z^{c,s}(\omega,\tau).$
Since the linear system we defined above is underdetermined, we cannot solve for $\vec{S}$ by straightforward linear inversion. Instead, we determine the prior mean through a regularized maximum likelihood estimate, 
\begin{align*}
\vec{S}_0\!=\!\text{argmin}_{\vec{S}} \,\frac{1}{2}(\vec{\mathcal{S}}_y^c\!-\!\mathbf{F}^c\vec{S})^T \mathbf{\Sigma}_c^{-1}(\vec{\mathcal{S}}_y^c\!-\!\mathbf{F}^c\vec{S})\!+\!|\!|\lambda \mathbf{D}(\vec{S}\!-\!\bar{S})|\!|^2,
\end{align*}
where $\mathbf{\Sigma}_c$ is the $9\times 9$ covariance matrix with elements $(\mathbf{\Sigma}_c)_{ss'}=\delta_{ss'}\text{var}[\hat{\mathcal{S}}_y^{c,s}]$. Note that the first term in this expression applies to a Gaussian likelihood function, valid in the limit of a large number of measurements. The rightmost term is an $L_2$ regularizer that ensures numerical stability \cite{TIKHONOV1963}, where $\lambda=0.35$ is the strength of the regularization, $\bar{S}$ is a constant $19\times 1$ vector containing the mean of the initial eigenestimates and $\mathbf{D}$ is a $19\times19$ diagonal matrix with nonzero elements, $D_{4,4}=\ldots=D_{10,10}=1$. The optimization has an analytic solution that yields the prior mean and corresponding covariance matrix,
\begin{align*}
&\vec{S}_0=(\mathbf{F}^{c\,T}\mathbf{\Sigma}_c^{-1}\mathbf{F}^{c}+2\lambda^2 \mathbf{D}^2)^{-1}\,
(\mathbf{F}^{c\,T}\mathbf{\Sigma}_c^{-1}\vec{\mathcal{S}}_y^c+2\lambda^2 \mathbf{D}^2\bar{S}),\\
&\mathbf{\Sigma}_0=(\mathbf{F}^{c\,T}\mathbf{\Sigma}_c^{-1}\mathbf{F}^{c}+2\lambda^2 \mathbf{D}^2)^{-1}.
\end{align*}
From these quantities, the prior distribution of $\vec{S}$ is
\begin{align}\label{eq::prior}
P(\vec{S})=\mathcal{N}_0\,e^{\frac{1}{2}(\vec{S}-\vec{S}_0)^T\Sigma_0^{-1}(\vec{S}-\vec{S}_0)},
\end{align}
where $\mathcal{N}_0$ is a normalization constant. Note that the prior is Gaussian-distributed since the $L_2$-regularizer preserves the Gaussianity of the likelihood function.

\begin{figure}[t]
\includegraphics[scale=.99]{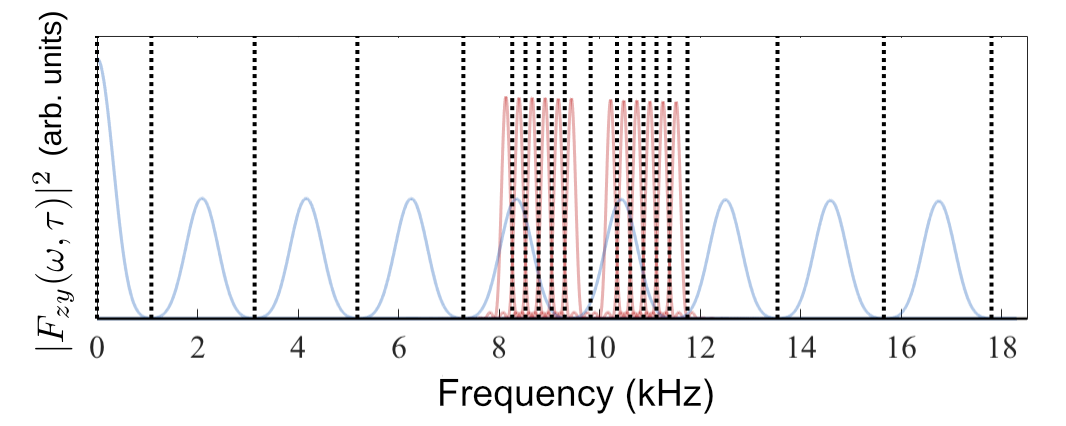}
\vspace*{-3mm}
\caption{Frequency segments used in Bayesian procedure. Boundaries of the frequency segments (dotted lines) are depicted along with the wide-band dephasing filters (blue) and narrowband dephasing filters (red).}
 \label{fig::segments}
\end{figure}

Next, we update the prior using data from additional measurements. To target the $8-12$ kHz region, the experiment used finite-difference sequences to generate 12 narrowband (or ``fine'', $f$) $k=0$ DPSS FFs. These FFs, centered at $\omega_s=2\pi f_s$ with $f_{s}= 8.1, 8.4,8.6,$ $8.9, 9.2, 9.4, 9.9, 10.2, 10.5, 10.9, 11.2, 11.5$ kHz, we denote by $F_z^{f,s}(\omega,\tau)$. In Fig. \ref{fig::segments} above, observe that each of the smaller segments contain a single narrowband FF, while each of the larger segments contain a single wide-band FF.  The measurements of $\mathcal{S}_y^{f,s}(\tau)\equiv\expect{|a_y^{(1)}(\tau)|^2}$ for each $F_z^{f,s}(\omega,\tau)$ we gather into a vector $\vec{\mathcal{S}}_y^f=[\hat{\mathcal{S}}_y^{f,1}(\tau),\ldots,\hat{\mathcal{S}}_y^{f,12}(\tau)]^T$. The corresponding $12\times 12$ covariance matrix has elements $(\Sigma_f)_{ss'}=\delta_{ss'}\text{var}[\hat{\mathcal{S}}_y^{f,s}]$.  Define now the $12\times19$ filter matrix $\mathbf{F}^f$ by $(\mathbf{F}^f)_{s,\ell}=\frac{1}{\pi}\int_{\sigma_\ell}\mathrm{d}\omega\,F_z^{f,s}(\omega,\tau)$, in analogy to $(\mathbf{F}^c)_{s,\ell}$ above.
The likelihood or conditional probability of measuring $\vec{\mathcal{S}}_y^f$ given the actual spectrum is then
\begin{align*}
P(\vec{\mathcal{S}}_y^f|\vec{S})=\mathcal{N}\,e^{\frac{1}{2}(\vec{\mathcal{S}}_y^f-\mathbf{F}^f\vec{S})^T\Sigma_{f}^{-1}(\vec{\mathcal{S}}_y^f-\mathbf{F}^f\vec{S})}.
\end{align*}
Again, we have assumed that each entry of $\vec{\mathcal{S}}_y^f$ is Gaussian distributed in the limit of a large  number of measurements. The posterior distribution is determined from the likelihood and prior in \erf{eq::prior},  
$P(\vec{S}|\vec{\mathcal{S}}_y^f)\propto P(\vec{\mathcal{S}}_y^f|\vec{S})P(\vec{S}).$
The mean of the posterior, which serves as our final spectral estimate, and the posterior covariance are given by
\begin{align*}
&\hat{\vec{S}}=(\mathbf{F}^{f\,T}\mathbf{\Sigma}_f^{-1}\mathbf{F}^{f}+\mathbf{\Sigma}_0^{-1})^{-1}\,
(\mathbf{F}^{f\,T}\mathbf{\Sigma}_f^{-1}\vec{\mathcal{S}}_y^f+\mathbf{\Sigma}_0^{-1}\vec{S}_0),\\
&\mathbf{\Sigma}=(\mathbf{F}^{f\,T}\mathbf{\Sigma}_f^{-1}\mathbf{F}^{f}+\mathbf{\Sigma}_0^{-1})^{-1}.
\end{align*}
The posterior mean, plotted in Fig. 2(b) of the main text, demonstrates improved resolution of the narrowband features in the 8-12 kHz region of the spectrum.

\begin{figure}[t]
\includegraphics[width=0.9\textwidth]{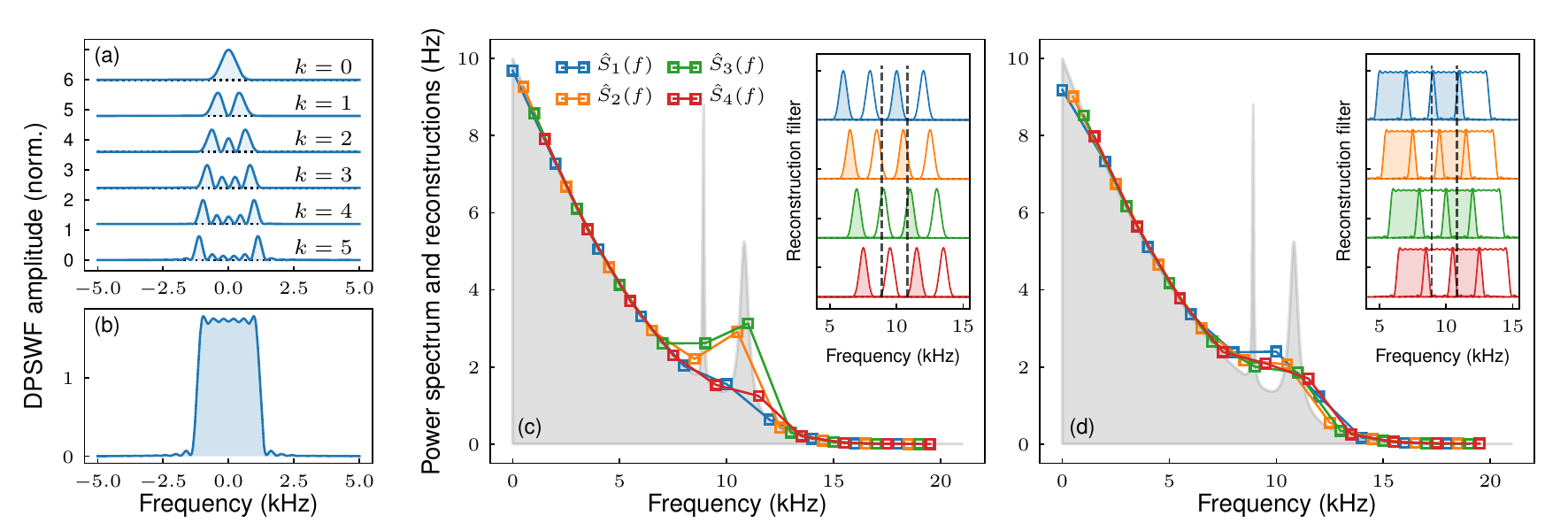}
\caption{Robust spectral estimation with finite-difference DPSS. \textbf{(a, b)} show different order DPSWF filters used for the spectrum reconstructions in panel \textbf{(c)} and \textbf{(d)}. The DPSS orders are in increasing order from $k=0$ to $k=5$. The sum of the filters (panel \textbf{(b)}) forms an approximate bandpass filter. \textbf{(c)} shows multiple single-taper reconstructions using the $k=0$ filters only whose centre frequencies are shifted by $\unit[+ i \times 0.5]{kHz}$ with respect to the first set of filters for $i \in \{0, 1, 2, 3\}$. This results in different reconstructions. \textbf{(d)} shows a multitaper reconstruction using the same filter centre frequencies. The inset shows the sum of DPSWF filters used for each reconstruction.}
\label{fig::multitaper}
\end{figure}

A key difference between the experimental procedure and the proposal outlined in Ref. \cite{Norris2018} is the use of $k=0$ DPSS FFs versus multitaper FFs in the initial detection stage. What we term multitaper FFs are actually composite FFs formed by the summation of measurements taken with Slepians of different orders \cite{Frey2017, Norris2018}. Let $F_z^{(k)}(\omega,\tau)$ be the finite-difference FF in \erf{eq::SuppFz}, corresponding to a Slepian of order $k$, and $\mathcal{S}_y^{(k)}(\tau)$ the resulting signal projection. For Slepians of orders $k=0,\ldots,K$, summing the signal projections weighted by coefficients $c_k$ produces
\begin{align*}
\sum_{k=0}^Kc_k\mathcal{S}_y^{(k)}(\tau)=\frac{1}{\pi}\int_{0}^\infty\!\!\!\!d\omega\,S_z(\omega)\bigg[\,\sum_{k=0}^Kc_kF_z^{(k)}(\omega,\tau)\bigg],
\end{align*}
where the quantity enclosed in the square brackets is the multitaper FF. With increasing $K$, the multitaper FF approaches an ideal bandpass filter, spectrally concentrated with uniform amplitude in its target band. In Fig. \ref{fig::multitaper}(c), the multitaper FF for $K=5$, $\omega_s=0$, and { $c_0=\ldots=c_5=1$}, 
%Virginia: are these weights correct?
is {\em spectrally concentrated and nearly uniform} within $(-2\pi W/\Delta t,2\pi W/\Delta t)$. This uniformity is an asset for detection, since a narrowband spectral feature will produce nearly the same signal at any location within the target band.  In contrast,  the $k=0$ Slepian FF is large in the center of the target band but falls to zero at the edges. As a result, a narrowband spectral feature near the edge of the band produces a substantially smaller signal projection than one near the center. The ability of $k=0$ Slepian FFs to detect narrowband spectral features, consequently, is highly dependent on the positions of the filters in general. In Fig. \ref{fig::multitaper}(d), this is illustrated with eigenestimates produced by four sets of $k=0$ Slepian FFs, each with slightly different positions along the frequency axis. The eigenestimates plotted in orange and green exhibit a large bump in the 8--12 kHz region since the second peak is positioned near the center of a band. The blue and red eigenestimates, on the other hand, register no bump since the second peak falls near the edge of a band. Thanks to their uniformity, multitaper FFs produce spectral estimates that are robust to position along the frequency axis. Figure  \ref{fig::multitaper}(e), shows estimates produced by sets of multitaper FFs with the same positions and bandwidth as the $k=0$ FFs. Unlike 
Fig. \ref{fig::multitaper}(d), each estimate exhibits a bump regardless of the FF positions.

Applying the multitaper estimation technique in experiments generally causes only a small overhead, as the number of measurements scale linearly with the number of included DPSS orders. However, experiments with engineered noise, as we have presented in the main text, require several random noise realizations to be implemented for each DPSS order ($\sim 400$ in Fig. 2 in the main text) to ensure that the spectrum is sampled uniformly and thus get an accurate representation for all tapers. For this reason, the experiments reported in the main text only used single-taper estimation, based on the $k=0$ DPSS order.

\section{Application to intrinsic system noise}
\label{app:intrinsic_noise}

\noindent We applied the multi-axis sensing protocol to probe the intrinsic noise in our system and to measure both native amplitude and dephasing noise over a range of $\unit[0-60]{kHz}$ with $\unit[20]{ms}$ long waveforms. The measurements, FFs and spectrum reconstructions are shown in Fig. \ref{fig::intrinsic_noise}. In our measurements we scaled the amplitudes of the DPSS waveforms quadratically with increasing band-shift frequency, such that the amplitude of the dephasing filters, which scales as $\propto 1/\omega^4$ [cf. \erf{eq::SuppFz}], remains constant under band-shifting. This comes at the cost of an unequal magnitude of the amplitude filters, which only scale as $\propto 1/\omega^2$,  and thus their amplitude increases with increasing band-shift.

The error vector components that we measured are well above our measurement fidelity limit of about 0.003, however we observed no clear frequency-dependent signature that could be attributed to the underlying noise processes. In the reconstruction of the amplitude quadrature in Fig. \ref{fig::intrinsic_noise}(f), we observe a $1/\omega^2$-type curvature which we believe may be an artifact of the scaling of the amplitude filters. We have separately performed a measurement (not shown) where the amplitude filters had their height held fixed, and we measured white noise at the $\unit[3\times10^{-7}]{Hz}$ level, which is consistent with the amplitude of last few reconstruction points in Fig. \ref{fig::intrinsic_noise}(f).  Variations in measurement outcomes have thus far made it impossible to definitively identify whether the observed behavior is a faithful representation of the ambient noise process.

The estimated dephasing noise spectrum shown in Fig. \ref{fig::intrinsic_noise}(g) is white over the measured frequency range. The extracted amplitude, however, appears to be approximately two orders of magnitude larger than that extracted from a Ramsey measurement (via the DC noise component).  $T_2$ echo measurements with CPMG waveforms yield similar results, which indicates that the DPSS dephasing filters may be sensitive to other noise processes in our experiment at such long interrogation times. For instance, we know that both intrinsic amplitude and dephasing noise are very low, and in fact our measured $T_2$ times are mostly limited by a reduction in ion fluorescence induced by ion-heating rather than pure dephasing noise. 

In summary, reconstructing intrinsic amplitude and dephasing noise spectra remains an ongoing project on our particular experimental platform, and more careful investigations are needed to determine potential error sources other than the two noise processes considered in our present theoretical framework. 

\begin{figure*}
\includegraphics[scale=0.9]{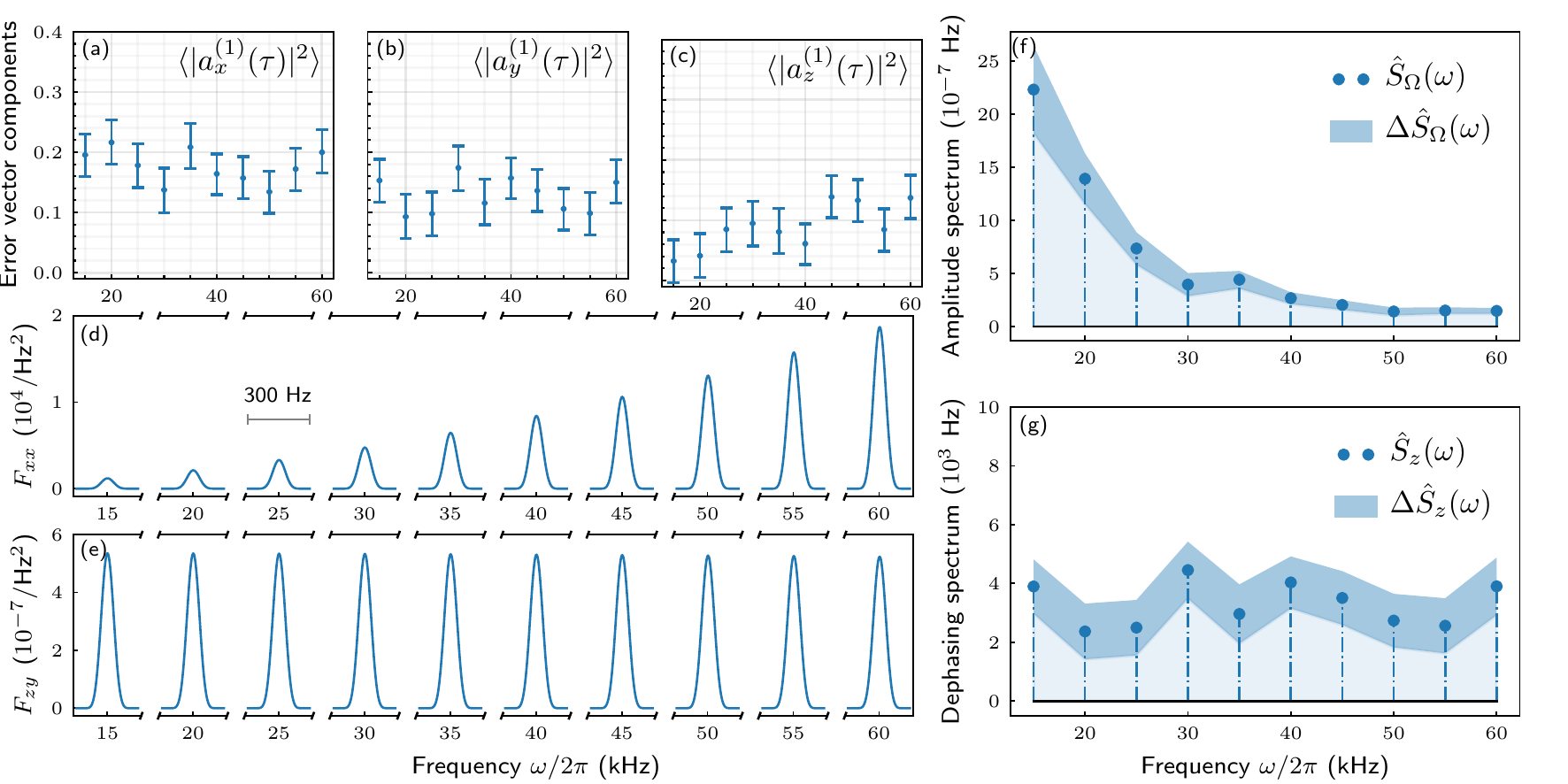}
\caption{Intrinsic noise spectral estimation. \textbf{(a, b, c)} show the measured error vector components [cf. Eqs. (\ref{ax}), (\ref{ay}) in the main text] taken with 10 band-shifted DPSS finite-difference waveforms with a duration of $\unit[20]{ms}$ sampled at $\unit[1]{MHz}$ with maximum Rabi rate $\Omega_\text{max}  \unit[\sim 15]{kHz}$. Each DPSS waveform was repeated 500 times and the error bars represent the standard deviations of the measurement outcomes. The corresponding amplitude and dephasing FFs are shown in \textbf{(d)} and \textbf{(e)} respectively, each plotted over a range of $\unit[300]{Hz}$. \textbf{(f)} shows the estimated amplitude spectrum at the reconstruction frequencies. The shaded area shows the estimated variance of the spectrum as calculated through propagation of the measurement errors. \textbf{(g)} shows the dephasing spectrum reconstruction.
}
\label{fig::intrinsic_noise}
\end{figure*}

\section{DPSS as discrete pulse sequences}

\noindent For experimental platforms which do not allow for arbitrary waveform generation, it is still possible to leverage the spectral concentration properties of DPSS through pulse sequences (see Fig. \ref{fig::dpss_style_sequences}). Consider a sequence of duration $\tau=N\Delta t$ consisting of $2N-1$ $\pi$-pulses about $\sigma_x$, applied at times $t_1,\ldots,t_{2N-1}$. The FF $|F_{zz}(\omega,\tau)|^2$, which arises in $\expect{|a_z^{(1)}(\tau)|^2}$ in \erf{eq::az2}, is spectrally concentrated about $\omega_s$ if the pulse times are chosen so that they depend on a COS modulated DPSS,
\begin{align}
\label{eq::PulsedSeq}
t_n=\begin{cases} n\Delta t/2,&n\;\text{even},\\
[c_\tau\,\cos(\frac{n-1}{2}\omega_s\Delta t)v_{\frac{n-1}{2}}^{(k)}(N,W)+n\Delta t]/2,&n\;\text{odd}.
\end{cases}
\end{align}
Here, $c_\tau$ is a scaling factor in units of time, satisfying $c_\tau\,v_n^{(k)}(N,W)<\Delta t$ for all $n$. A sample sequence generating a $k=0$ DPSS FF centered at $\omega_s=0$ is illustrated in Fig. \ref{fig::dpss_style_sequences} (a)-(c). Experimental reconstructions of such a FF centered at $\omega_s=0$ is shown in \ref{fig::dpss_style_sequences} (d). The signal measured at $\unit[\sim 530]{Hz}$ is the harmonic of the sequence generated by the periodic spacing of the $\pi$-pulses at the end of each segment (green pulses in Fig. \ref{fig::dpss_style_sequences}(b)). If the target spectrum to be estimated has a very wide frequency range, care must be taken in ensuring that leakage bias resulting from that may be accounted for. 

\begin{figure}[t]
\includegraphics[width=0.9\textwidth]{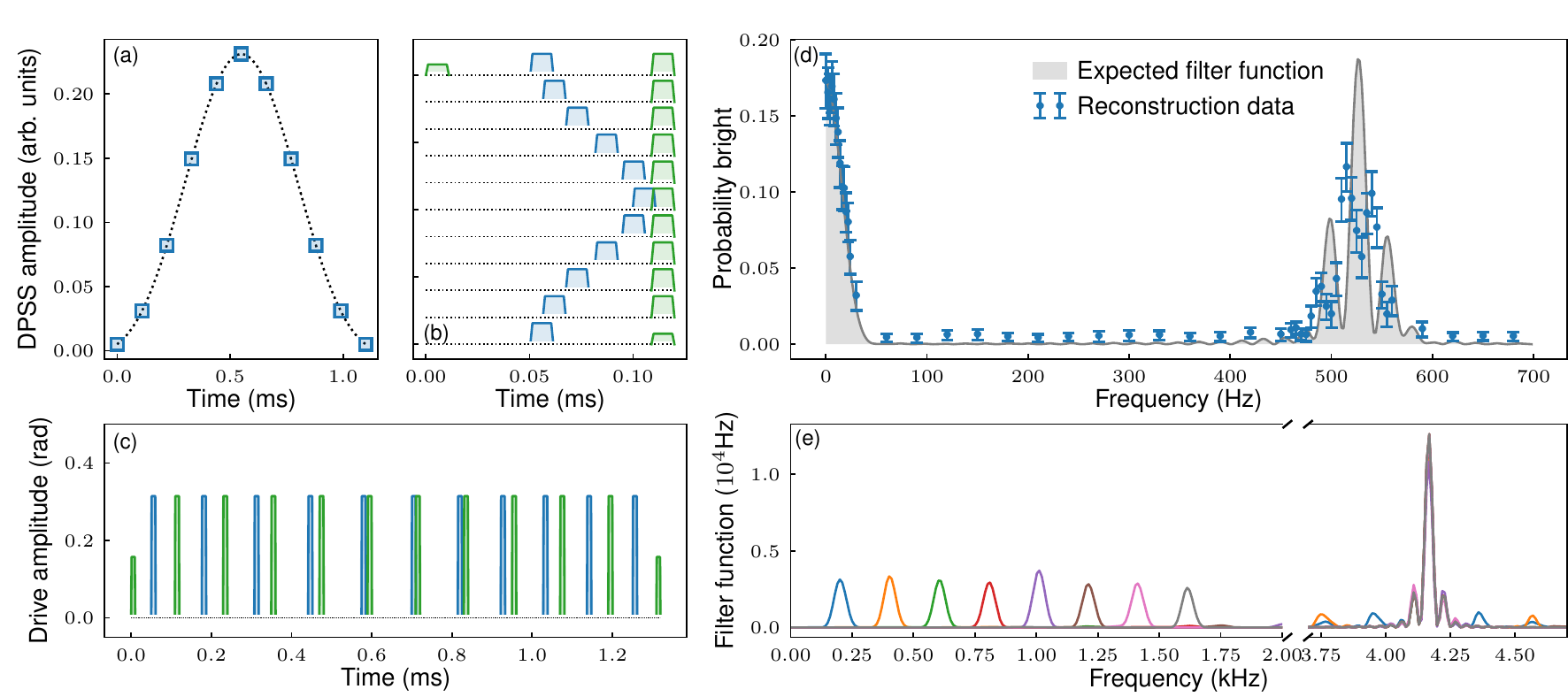}
\caption{Discrete DPSS pulse sequences. \textbf{(a)} shows a sample DPSS which informs the pulse locations in the pulse sequence in panel \textbf{(b)}. The discrete points in time of the DPSS sequence in \textbf{a} are marked by the blue squares and correspond to the blue $\pi$-pulses in panel \textbf{(b)}. The green $\pi$-pulses mark the end of a segment in the sequence. The full, ravelled sequence is shown in \textbf{(c)} with the same colour coding. \textbf{(d)} shows an experimental FF reconstruction of a $k=0$ DPSS pulse sequence with 25 segments with a $\pi$-time of $T_\pi = \unit[35]{\upmu s}$ and additional free evolution time of about $\unit[1.8]{ms}$ per segment, making the whole sequence $\unit[50]{ms}$ long. The FF of the corresponding sequence is mapped out by supplying a single-tone modulation to the FM input of the signal generator, and scanning the frequency of that modulation. \textbf{(e)} shows numerically calculated FFs of a series of bandshifted DPSS pulse sequences with 50 segments of $\unit[220]{\upmu s}$ duration and with a $\pi$-time of $T_\pi = \unit[10]{\upmu s}$. The main harmonic at $1/(\unit[220]{\upmu s} + \unit[2 \times 10]{\upmu s}) \approx \unit[4.166]{kHz})$ is the same for all filters.}
\label{fig::dpss_style_sequences}
\end{figure}

To better understand the spectral concentration properties of the FFs generated by these pulse sequences, consider the idealized case in which the $\pi$-pulses are instantaneous and the amplitude control waveform takes the form
%\begin{align*}
$\Omega(t)=\pi\sum_{n=1}^{2N-1}\delta(t-t_n).$
%\end{align*}
For this amplitude waveform, $\cos\Theta(t)\mapsto y(t)$ in \erf{eq::az2}, where $y(t)$ is a switching function that toggles between $\pm 1$ with every application of a $\pi$-pulse. More concretely, the switching function takes the values
\begin{align*}
y(t)=\begin{cases} 1&0\leq t<t_1,\\
-1&t_1\leq t<t_2,\\
1&t_2\leq t<t_3,\\
%-1&t_3\leq t<t_4,\\
\quad\quad\vdots\\
1&t_{2N-2}\leq t<t_{2N-1},\\
-1&t_{2N-1}\leq t<\tau.
\end{cases}
\end{align*}
The Fourier transform is then
\begin{align*}
F_{zz}(\omega,\tau)=\int_0^\tau\!\!d\mathrm{s}\,e^{i\omega s}y(s)
=\sum_{m=0}^{N-1}e^{i\omega m\Delta t}\int_0^{\Delta t}\!\!d\mathrm{s}\,e^{i\omega s}y(s+m\Delta t).
\end{align*}
If $\omega_c$ is the approximate cutoff frequency of the spectrum, i.e., $S_z(\omega)\approx 0$ for $\omega>\omega_c$, and $\omega_c\Delta t\ll 1$, then $e^{i\omega m\Delta t}\approx 1$ in the above expression. From \erf{eq::PulsedSeq}, letting $n'=2m$,
\begin{align*}
\int_0^{\Delta t}\!\!\mathrm{d}s\,y(s+m\Delta t)=&\int_{t_{n'}}^{t_{n'+1}}\!\!\mathrm{d}s-\int_{t_{n'+1}}^{t_{n'+2}}\!\!\mathrm{d}{s} 
= c_\tau\cos(m\omega_s\Delta t)v_m^{(k)}(N,W).
\end{align*}
The FF entering \erf{eq::az2} is then
\begin{align*}
|F_{zz}(\omega,\tau)|^2\approx\,\bigg|\sum_{m=0}^{N-1}e^{i\omega m\Delta t}c_\tau\cos(m\omega_s\Delta t)v_m^{(k)}(N,W)\bigg|^2
\approx c_\tau^2\Big[\,\,|U^{(k)}(N,W;\omega-\omega_s)|^2+|U^{(k)}(N,W;\omega+\omega_s)|^2\,\Big],
\end{align*}
which is spectrally concentrated about $\omega_s$, as desired.

\end{widetext}

\end{document}